\documentclass[draft]{agujournal2019}
\usepackage{url}
\usepackage{lineno}
\usepackage[inline]{trackchanges} 
\usepackage{soul}
\usepackage{acronym}
\usepackage{amsmath}
\acrodef{GLE}[GLE]{Ground Level Enhancement}
\acrodef{SC}[SC]{solar cycle}
\acrodef{SEP}[SEP]{Solar Energetic Particle}
\acrodef{HEPAD}[HEPAD]{High Energy Proton and Alpha Detector}
\acrodef{NM}[NM]{Neutron Monitor}
\acrodef{FD}[FD]{Forbush Decrease}
\acrodef{CME}[CME]{Coronal Mass Ejection}
\acrodef{GCR}[GCR]{Galactic Cosmic Ray}
\acrodef{AATB}[AATB]{Almaty NM, Kazakhstan}
\acrodef{HRMS}[HRMS]{Hermanus Neutron Monitor}
\acrodef{FSMT}[FSMT]{Fort Smith Neutron Monitor}
\acrodef{SOPO}[SOPO]{South Pole Neutron Monitor}
\acrodef{DOMB}[DOMB]{Dome C, Central Antarctica Neutron Monitor}
\acrodef{OULU}[OULU]{Oulu, Finnland Neutron Monitor}
\acrodef{THUL}[THUL]{Thule neutron monitor}
\acrodef{IRK3}[IRK3]{Irkutsk3 Neutron Monitor}
\acrodef{MT}[MT]{Muon Telescope}
\acrodef{DOSTEL}[DOSTEL]{DOSimetry TELescope}
\acrodef{ISS}[ISS]{International Space Station}
\acrodef{LEO}[LEO]{Low Earth Orbit}
\acrodef{AMS}[AMS]{Alpha Magnet Spectrometer}
\acrodef{IGRF}[IGRF]{International Reference Geomagnetic Field}
\acrodef{TSY89}[TSY89]{Tsyganenko ’89}
\acrodef{TSY96}[TSY96]{Tsyganenko ’96}
\acrodef{TSY01}[TSY01]{Tsyganenko ’01}
\acrodef{TSY05}[TSY05]{Tsyganenko ’05}
\acrodef{ESA}[ESA]{European Space Agency}
\acrodef{CAU}[CAU]{Christian-Albrechts-University}
\acrodef{DLR}[DLR]{Deutsches Zentrum für Luft und Raumfahrt}
\acrodef{PIPS}[PIPS]{Passivated Implanted Planar Silicon}
\acrodef{SAA}[SAA]{South Atlantic Anomaly}
\acrodef{LET}[LET]{Linear Energy Transfer}
\acrodef{EPHIN}[EPHIN]{Electron Proton Helium INstrument}
\acrodef{SOHO}[SOHO]{SOlar and Heliospheric Observatory}
\acrodef{DOY}[DOY]{Day of Year}

\newcommand{\apjl}{Astrophysical Journal Letters}

\journalname{Space Weather}

\begin{document}

\title{Yield function of the \acl{DOSTEL} count and dose rates aboard the \acl{ISS}}

\authors{Caprotti, A. S.\affil{1,}\affil{2}, Br\"udern, M.\affil{1}, Burmeister, S.\affil{1}, Heber, B.\affil{1}, and Herbst, K.\affil{1}}

\affiliation{1}{Institut f\"ur Experimentelle und Angewandte Physik, Christian-Albrechts Universit\"at zu Kiel,
Leibnizstra\ss e 11, D-24118 Kiel, Germany}
\affiliation{2}{Università degli Studi di Milano, Dipartimento di Fisica, Via Celoria 16, 20133 Milano, Italy}

\correspondingauthor{Heber, B.}{heber@physik.uni-kiel.de}

\begin{keypoints}
\item Energetic Particles
\item Influence of the Earth magnetosphere on the count- and dose rate aboard \ac{ISS} 
\item Yield function
\end{keypoints}

\begin{abstract}
The Earth is constantly hit by energetic particles originating from galactic sources. The flux of these particles is altered by the magnetized solar wind in the heliosphere and the Earth's magnetic field. For this reason, the ability of a particle to approach a spacecraft in \ac{LEO} depends on its energy and the  position of the spacecraft within the Earth' magnetosphere. Moreover, there are some areas (radiation belts) where the particles are trapped for a long time, and therefore the flux of energetic particles is particularly high. Occasionally, \acp{SEP} contribute to the energetic particle flux too. \ac{DOSTEL} is one of the instruments aboard the \ac{ISS} that monitors the radiation field within the European module Columbus. Because being installed inside the \ac{ISS}, particles produced by the interaction between the ''primary'' radiation and the \ac{ISS} materials are also measured. To describe the observations in such a complex radiation field, we follow the method by \citeauthor{Caballero-Lopez-Moraal-2012} \citeyear{Caballero-Lopez-Moraal-2012} in order to compute the so-called yield function using precise measurements of the proton and Helium energy spectra obtained by \ac{AMS} and the systematic variation of the \ac{DOSTEL} measurements within the Earth's magnetosphere. 
\end{abstract}

\section{Introduction}
The radiation environment close to the Earth is dominated by energetic charged particles covering the energy range from below a few keV to about $10^{21}$ eV. At altitudes of about 400~km and with an orbital inclination of 51.6 degrees, both the magnetic field and the hull of the ISS shield against the lowest energies. The origin of the particles that contribute most to the radiation dose is of galactic and solar or comes from the captured particles within the radiation belts \cite{Xapsos-etal-2013}. 
These trapped particles (primarily protons) are measured during passages of the so-called \ac{SAA}. Of special interest is the radiation field within the \ac{ISS}. Due to the interaction of energetic particles with the \ac{ISS} material this field differs significantly from the one outside and from the position of the station and within the station (see for example \citeauthor{Labrenz-etal-2015} \citeyear{Labrenz-etal-2015} or \citeauthor{Berger-etal-2017} \citeyear{Berger-etal-2017}). However,  the shielding is much lower than the one at sea-level and might be compared to the radiation environment in the lower Earth stratosphere \cite{Caballero-Lopez-Moraal-2012}. In addition to the  shielding, the particle flux is altered by the Earth's magnetic field that can be approximated by a tilted dipole, which has an offset with respect to the Earth's center. As detailed below, different mathematical models exist to describe its geometry. As a consequence of this, one measures energetic particle fluxes at \ac{ISS} altitudes that depend on the geomagnetic position \cite[and references therein]{Labrenz-etal-2015}.  During \acp{GLE} \acp{SEP} are measured by instruments aboard the \ac{ISS} \cite[and references therein]{Berger-etal-2018}. In space physics research, the yield function of a detector (i.e., neutron monitor, muon telescope, etc.) is defined as the relationship between the intensity of primary cosmic rays and the counting rate of an instrument inside the atmosphere. As detailed below, the count rate depends on the geomagnetic position characterized by the so-called cutoff rigidity (see Eq.~\ref{eq:yield-def}). There are two ways to determine the atmospheric yield function. The empirical method is to measure the counting rate of an instrument inside simultaneously, and the primary spectrum outside the atmosphere \cite[and references therein]{Aiemsa-ad-etal-2015, Mangeard-etal-2016a}. The other method is to calculate the yield function by using one of several generic numerical codes such as, {for example, FLUKA \cite{Bohlen-etal-2014} and GEANT4 \cite{Agostinelli-etal-2003} } to simulate the atmospheric cascade process, and then adding detector-specific details to these models  \cite[respectively]{Mangeard-etal-2016b, Mishev-etal-2020}. 

When particles interact with human bodies and materials in the spacecraft, they deposit energy in the material. The energy per unit mass of the target is a quantity called dose (units: Gy). DOSIS 3D is an experiment aiming to study the dose distribution within the European Columbus module of the \ac{ISS}. The active component of this experiment is the \ac{DOSTEL} that measures the count and dose rates as a function of time. For more detail on the instrument and the measurements onboard the ISS see \cite{Labrenz-etal-2015}. In order to describe the response function in such a complex radiation field, we follow the approach detailed in \citeauthor{Caballero-Lopez-Moraal-2012} \citeyear{Caballero-Lopez-Moraal-2012} and determine the  \textit{yield function} empirically using precise measurements of the proton and Helium energy spectra outside the Earth and the systematic variation of the \ac{DOSTEL} measurements inside the \ac{ISS} as a function of the cutoff rigidity. In what follows, we describe the instrument, the measurements utilized in our study, and the theoretical background. The yield function is then determined during two quiet periods that are times when the solar and Earth magnetic field activities are low. 
\section{Instrumentation}
In 2012, the DOSIS3D experiment developed by \ac{CAU}, Kiel (Germany) and \ac{DLR}, Köln (Germany) was installed inside the European Columbus Laboratory on board of the \ac{ISS}. One of the main goals is to measure radiation exposure inside the ISS to estimate radiation risks for future missions. The instrument setup is composed of passive and active detectors:  The first ones measure integrated values of the dose received during their exposure; the others provide real-time information. Each active detector is a \ac{DOSTEL}. A previous version has been applied on-board several Space Shuttle missions of the \ac{ESA} in the year 1996-1997 \cite{Beaujean-etal-2002,Singleterry-etal-2001} and mounted on the MIR. In 2001 the instrument was mounted in the US Laboratory on the ISS as part of Dosimetry Mapping (DosMap), the first European Dosimetry experiment \cite{Reitz-etal-2009}. Between 2004 and 2011, a further version was used in the MATROSHKA experiment \cite{Labrenz-etal-2015}.

\begin{figure}
    \centering
    \includegraphics[width=\columnwidth]{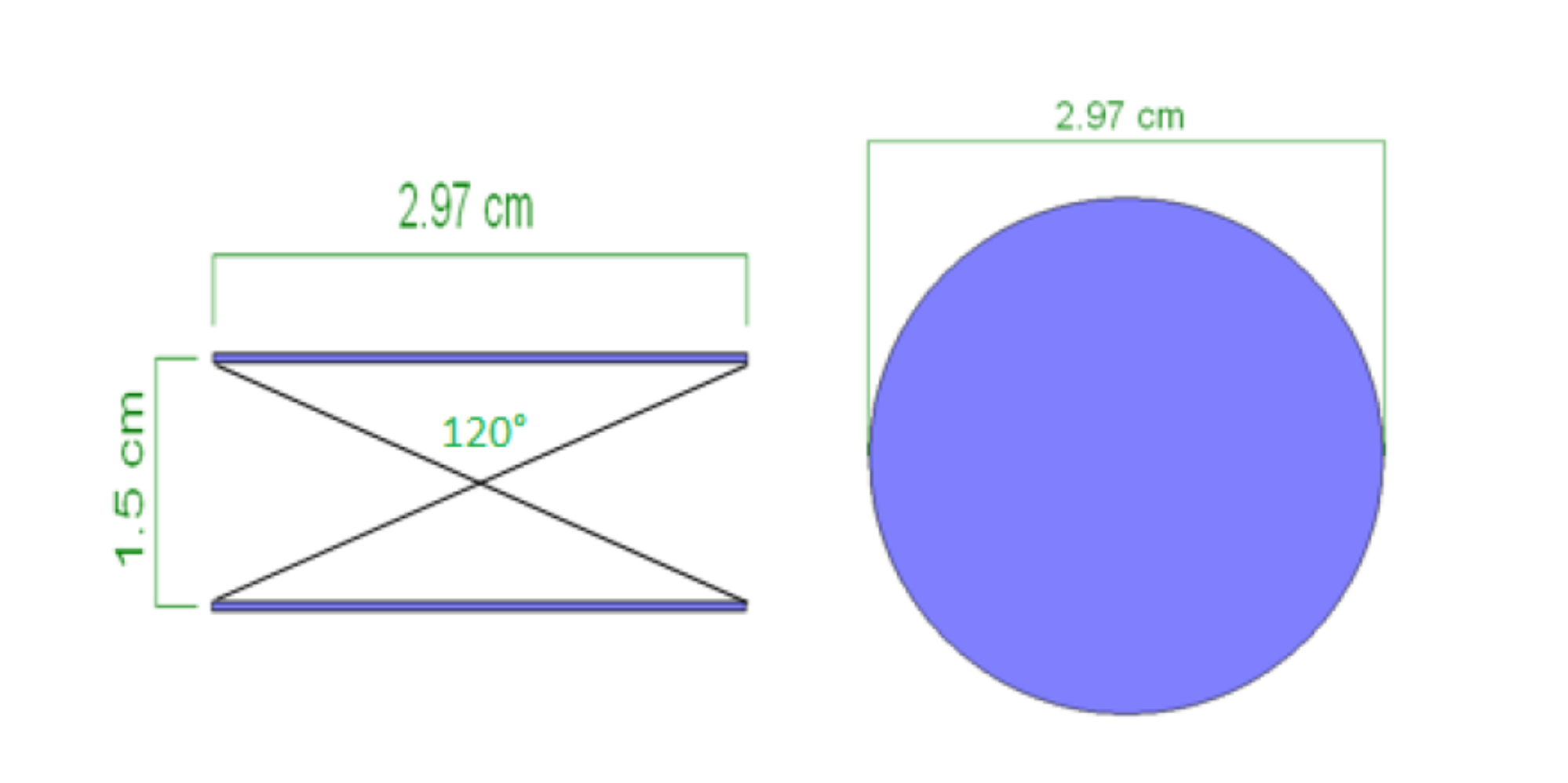}
    \caption{Sketches from side (left) and top (right) of a \ac{DOSTEL}}
    \label{fig:DOSTEL-SCHEME}
\end{figure}

Each \ac{DOSTEL} consists of two \ac{PIPS} detectors, each with a thickness of 315~$\mu$m and an area of 6.93~cm$^2$ arranged in a telescope geometry (see Fig.~\ref{fig:DOSTEL-SCHEME}). The distance between the two detectors is 1.5~cm. Both the opening angle and the geometric factor of the instrument are 120$^\circ$ and 824~mm$^2$ sr for particles in coincidence mode (i.e., hitting both the detectors: these are called “telescope” or \ac{LET} measurements). {The instrument measures count rates and dose rates of radiation hitting a single detector (“dose measurement”). Each detector is sensitive to ions and electrons including minimum ionizing particles, photons in the energy range above the energy threshold of the detector as well as neutrons \cite{Moeller-2008}.  } Focusing on the dose measurements, the particle rate and the absorbed dose rate are stored in the \ac{DOSTEL} memory after a certain time interval. In order to have a good statistic, this time-interval is set to 100~s outside the region of the \ac{SAA}, where the count rates are less than 30~per second. In the \ac{SAA}, the time interval is chosen to be 20~s to improve time resolution. From these measurements, the absorbed dose rates are calculated. {In contrast}, the integration time for data in coincidence mode is about 45~minutes, and corresponding data sets of are stored separately inside and outside the \ac{SAA} as histograms of deposited energy. { From the energy deposition spectra LET spectra are derived to obtain the mean quality factor according the Q(L) dependence given in ICRP60 \cite{ICRP60} as:
\begin{linenomath*}
\begin{equation}
Q(L)=\begin{cases}
1 & \text{for } L < 10~\textrm{keV}/\mu\textrm{m}\\
0.32 L - 2.2&  \text{for }~10\le L \le 100~\textrm{keV}/\mu\textrm{m}\\
\frac{300}{\sqrt{L}} & \text{for } L > 100~\textrm{keV}/\mu\textrm{m}
\end{cases}
\end{equation}
\end{linenomath*}
Dose equivalent can be calculated by multiplying the mean quality factor to the measured dose:}
Due to the limited angle of incidence the mean path length of 364 $\mu$m in silicon is used to obtain the dE/dx in silicon. To convert dE/dx in silicon into dE/dx in water the different stopping power of high energetic charged particles was used to calculate a mean conversion factor of 1.23. This approximation for the LET was concluded to be sufficient for dosimetry purposes. Since two \ac{DOSTEL} instruments are mounted perpendicular to each other, information about the directionality of the radiation field inside the Columbus module can be determined \cite{Berger-etal-2017}.

\section{Observations}
\begin{figure}
    \centering
    \includegraphics[width=\columnwidth]{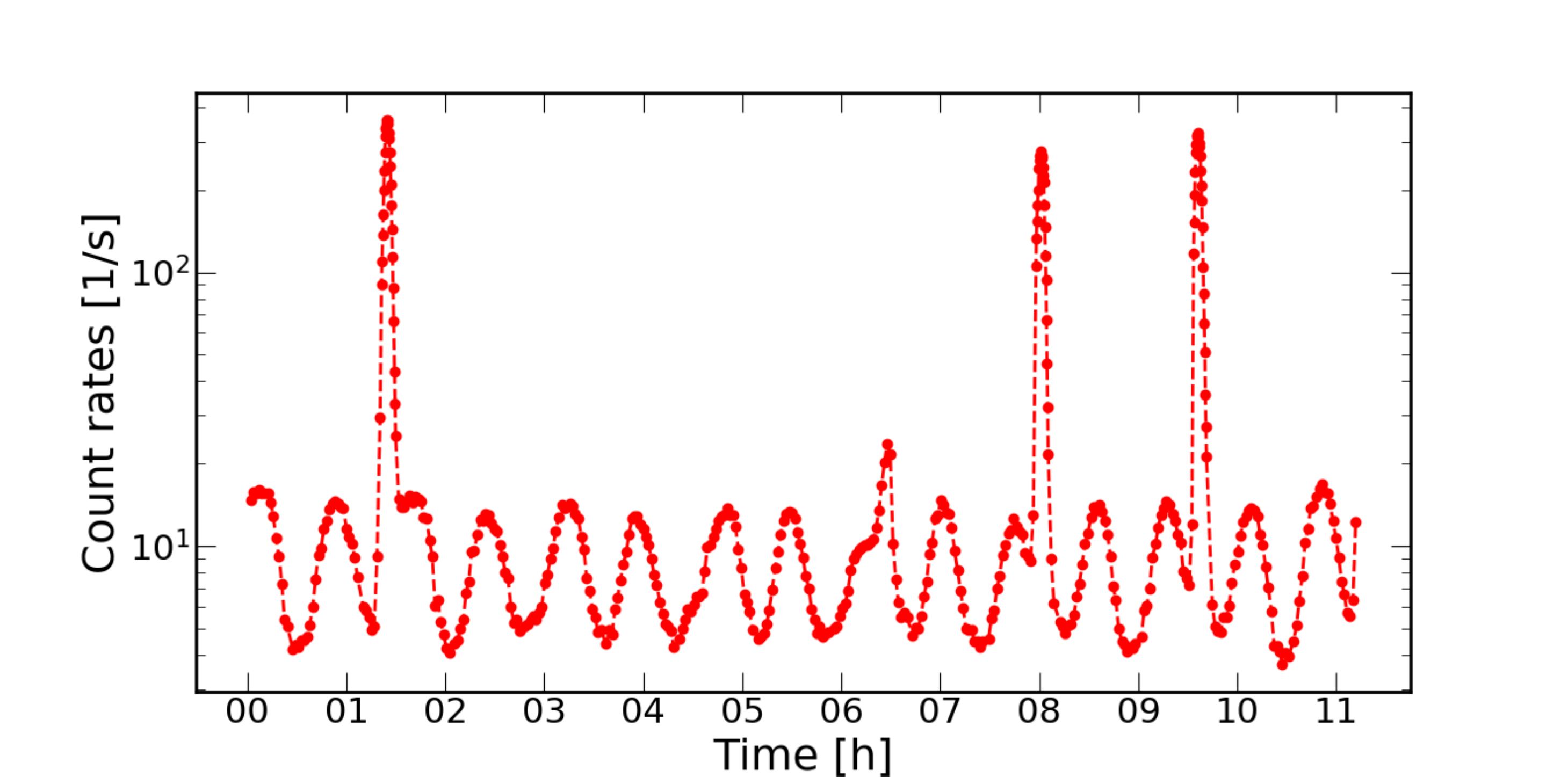}
    \caption{Count rates measured by \ac{DOSTEL} for a sample of 11 hours on 1/1/2014.The sudden count rate increases around 1:25~h, 6:28~h, 8:01~h, and 9:36~h correspond to crossings of the \ac{SAA}, respectively. }
    \label{fig:DOSTEL-RATES}
\end{figure}
Exemplarily, Fig.~\ref{fig:DOSTEL-RATES} displays the observed count rate profile of the first half of January 1, 2014. Thereby, high and low count rates correspond to high and low latitudes, respectively. The sudden count rate increases around 1:25~h, 6:28~h, 8:01~h, and 9:36~h correspond to crossings of the \ac{SAA}, respectively. In the following analysis, these data points are excluded.

The variation of the count rates depends significantly on the geomagnetic cutoff rigidity $R_C$.  Thus, the count rate $N$ is a function of $R_C$ and is given by: 
\begin{equation}\label{eq:yield-def}
    N(R_C,t) = \sum\limits_{i} \int\limits_{R_C}^{\infty} dR \; j_i(R,t) \; Y_i(R),
\end{equation}
where $j_i(R,t)$ is the flux of incoming particles of the species $i$ (Proton, Helium, Z$>$3) and rigidity $R$ at time $t$, and $Y_i(R)$ is the species dependent yield function of the \ac{DOSTEL} instrument. The yield function depends on the rigidity, because particles with higher rigidities are more likely to produce a large number of secondary particles. For the determination of the yield function we follow the approach that was suggested by \citeauthor{Caballero-Lopez-Moraal-2012} \citeyear{Caballero-Lopez-Moraal-2012}. To do so we had to perform the following tasks: 
\begin{enumerate}
\item The cutoff rigidity dependency of the count rate $N(R_C)$ is approximated by the Dorman function \cite{Dorman-etal-1970}:
    \begin{equation}
        N(R_C) = N_0 \; \left[ 1-\exp\left(-\alpha R_C^\kappa\right)\right],
    \label{eq:DORMAN}
    \end{equation}
with $\alpha$ and $\kappa$ as free parameters depending on the altitude of the \ac{ISS} and the phase in the solar cycle, respectively
\item The weight function $F_i(R)$ of each species $i$ to an averaged yield function $Y(R)$ is determined, 
\item the precise knowledge of the primary rigidity spectra $j_i(R)$ of each species utilizing  flux data from \ac{AMS} published by \citeauthor{Aguilar-etal-2018} \citeyear{Aguilar-etal-2018}  
\end{enumerate}
In the following sections, each of these required steps is discussed.
\section{Theoretical background}
As mentioned above, \ac{DOSTEL} data have to be analyzed to {find} the yield function of the instrument. The calculation requires a good description of the measured count rates within the Earth's magnetosphere by Eq.~(\ref{eq:DORMAN}), which further relates the \ac{DOSTEL} count rates to cutoff rigidities. Note that the latter is computed from the geographical positions of the \ac{ISS} obtained by great circle interpolations using \ac{ISS} orbital data.  Thus, the first step is to compute $R_C$ using a certain model of the Earth's magnetic field (see Sec.~\ref{sec:Earth-Magnetic_Field}) utilizing the PLANETOCOSMICS code described in \citeauthor{Desorgher-etal-2009} \citeyear{Desorgher-etal-2009} (see Sec.~\ref{sec:PLANETOCSOMICS}). In order to facilitate the computation, some approximations have to be introduced, and their influence has to be investigated. 
\subsection{The Earth's magnetic field} \label{sec:Earth-Magnetic_Field}
Within a distance of about 6~R$_{E}$ from the Earth's surface, the internal geomagnetic {field} can be approximated as a dipole, tilted with respect to the Earth's spinning axis by an angle of about $11^\circ$. Moreover, the center of the dipole is nearly 400~km {far-away} from the Earth's center \cite{Walt-1994}. For distances further out ($>$ 6~R$_{E}$) the magnetic {field} shape is significantly affected by the pressure of the solar wind: the {field} lines are squeezed on the dayside towards the Sun, and they cover a region within $\sim$~10~R$_{E}$ from the Earth’s surface. On the nightside, the lines are stretched, and they extend up to $\geq$~50~R$_{E}$. 

Semi-empirical models can describe the internal and external {field}s. In this study, we used the \ac{IGRF} model \cite[in particular IGRF-12]{Thebault-etal-2015} as a representation of the internal magnetic field, and the \ac{TSY89} model representing the external magnetic field. The planetary magnetic field disturbance level is quantified by the so-called global {$K_p$} index, derived via experimental procedures. Every three hours, at 13 ground-based magnetic observatories in subauroral regions, variations of the horizontal components of the magnetic {field} are measured. Then, the variation range of each component (the difference between the highest and the lowest values) is considered. The measure of the largest range is associated with a certain value (between 0 and 9) of a local {$K_p$} index. As summarized in \citeauthor{Tsyganenko-2013} \citeyear{Tsyganenko-2013} the \ac{TSY89} model has further been improved (\ac{TSY96}, \ac{TSY01}, and recently \ac{TSY05}). Nevertheless, as discussed in \citeauthor{Tsyganenko-2013} \citeyear{Tsyganenko-2013} the model results based on the \ac{TSY89} and the most recent \ac{TSY05} model show reasonable agreement for k$_p < 3$ conditions. Therefore, in this study, the \ac{TSY89} model was chosen to perform the simulations for the sake of simplicity, only requiring the {$K_p$} values as an input parameter. We note that only low {$K_p$} conditions {($k_p\leq 4$)} are considered.
\subsection{Computation of Vertical Cutoff Rigidities} \label{sec:PLANETOCSOMICS}
\begin{figure}
   \centering
    \includegraphics[width=\columnwidth]{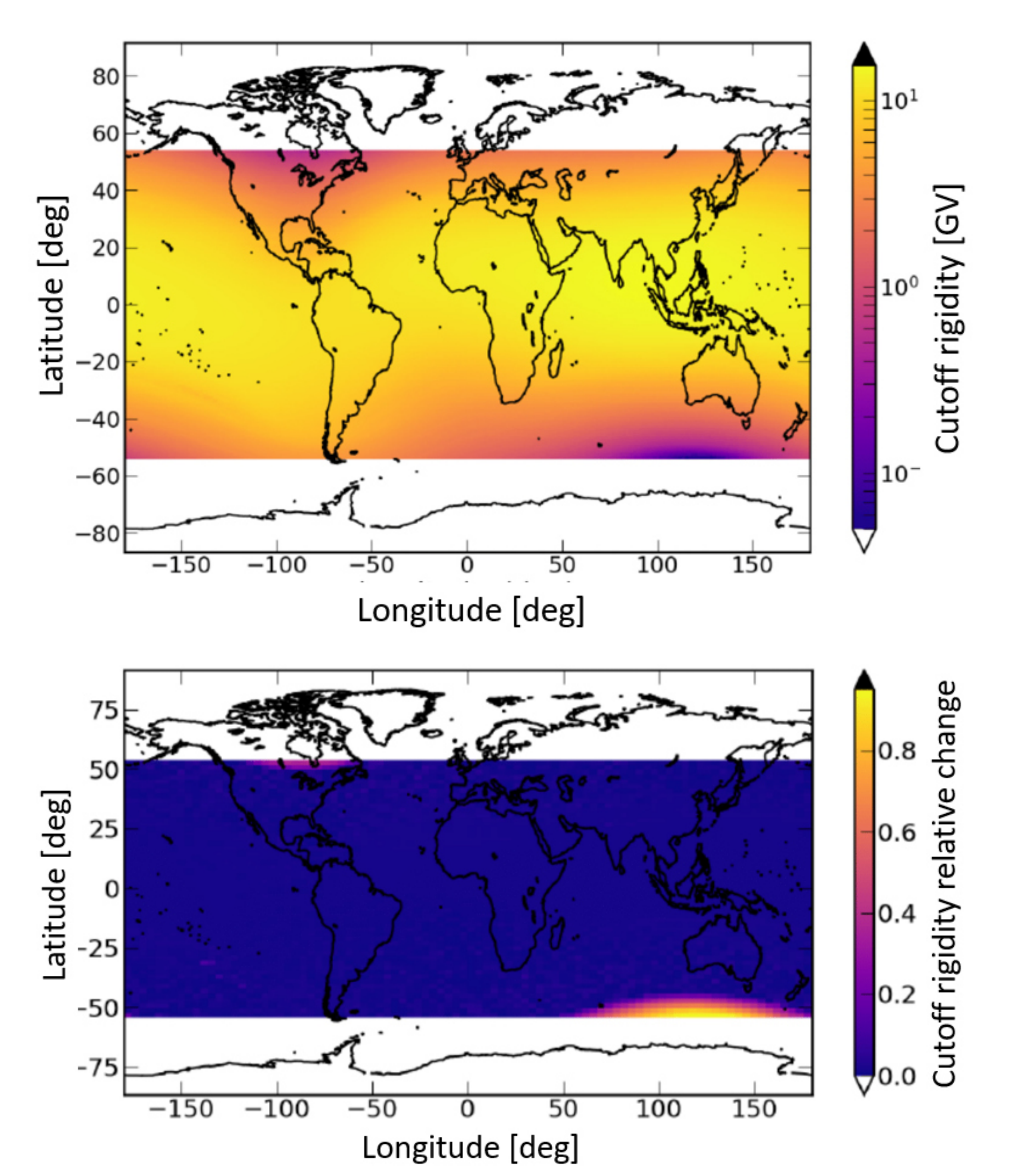}
    \caption{Upper panel: Vertical cutoff rigidities simulated with PLANETOCOSMICS for March 15, 2013 at an altitude of 415~km.  The figure shows that the values decrease from the Equator to the Poles. 
    Lower panel: The relative difference in cutoff rigidity computed on March 13, 2013 at the minumum and maximum altitude of the \ac{ISS} (399~km and 440~km, respectively).}
    \label{fig:Figure-3-4}
    \label{fig:PLANETOCOSMICS-MAP}
\end{figure}

The motion of charged particles in magnetic fields is described by the particle rigidity \cite{Shea+Smart-1965, Desorgher-etal-2009}. Although depending on the angle between the velocity of the particle and the magnetic field (pitch angle), we characterize the accessibility by the so-called vertical cutoff rigidity, the rigidity a particle has being measured by an upward pointing detector. Note, it is more difficult for charged particles in general to reach low latitudes than higher latitudes. Fig.~\ref{fig:PLANETOCOSMICS-MAP} shows the global distribution of the vertical cutoff rigidities modeled at an \ac{ISS} altitude of 415 km. Different colors correspond to different cutoff rigidity values. Depending on cutoff rigidities, the count rates for a detector at a {fixed} altitude change with the geographical coordinates. This is illustrated by  \citeauthor{Picozza-etal-2013} \citeyear{Picozza-etal-2013} in their Fig.~13 showing PAMELA measurements at different laltitudes. 
Their measurements show a steep intensity decrease at the rigidities below $R_c$.  
To derive these cutoff rigidity maps, we utilized the PLANETOCOSMICS code described in \citeauthor{Desorgher-etal-2009} \citeyear{Desorgher-etal-2009}: particles are generated at different rigidities at a given position at the altitude of the \ac{ISS} and a{n} incidence radial towards the center of the Earth (vertical direction). Computing their propagation in the geomagnetic {field}, for each particle a particle with opposite charge is chosen that is injected at the given position and its direction opposite to the incoming direction \cite{McCracken+Freon-1962}. This is known as the ''backward-trajectory tracing method'' and described in Sec.~\ref{sec:Earth-Magnetic_Field}. Note, that in PLANETOCOSMICS not the planetary index {$K_p$} is used but a modified version of it, the option parameter IOPT (see Tab.~\ref{tab:KP-IOPT}) following the definition by \citeauthor{Tsyganenko-1989}, \citeyear{Tsyganenko-1989} (see also \citeauthor{Kudela-Usoskin-2004}\citeyear{Kudela-Usoskin-2004}).
\paragraph*{Altitude dependence}
Within 2013, the ISS maneuvered at altitudes between 399.322 km and 439.865 km above the terrestrial surface. To estimate the uncertainties due {to} the altitude variations, we calculated the cutoff rigidities for specific times and IOPTs, varying the altitudes among five chosen values. The starting time of the simulation was set to 15.00~UTC on June 27, 2014, with an IOPT value of 2 for the upcoming three hours. The lower panel of Fig.~\ref{fig:Figure-3-4} displays the relative differences in the computed cutoff rigidity values for an altitude of 399~km and 440~km corresponding to the lowest and highest orbit altitude of the ISS, respectively. This is a good representation, since the ISS orbit varied only between 415 km and 430 km in 2013. The corresponding relative differences range between 0 and nearly 1.

The differences increase with increasing latitude. A possible reason could be the following: near the Equator, even at different altitudes, the {field} lines crossed by the \ac{ISS} are almost parallel to the Earth’s surface and therefore to each other; going towards the poles, the {field} lines move closer to each other, and therefore they are not parallel anymore. The comparisons between the simulations revealed the highest relative error to be around  {0.8 GV} in a few bins at the highest latitudes considered. Comparing the lower panel of Fig.~\ref{fig:Figure-3-4} with the upper panel, we find that these regions correspond to cutoff rigidities up to some tenth of GV. In this study, only particles with a rigidity above $\sim$0.5~GV are taken into account. For these particles, the altitude-dependent differences, however, are seldom that high. Moreover, the comparison is made for two situations with a very high altitude difference, which likely leads to an overestimation of the relative error.
\paragraph*{Variation of the IOPT parameter in PLANETOCOSMICS:} \begin{figure}
\centering
    \includegraphics[width=\columnwidth]{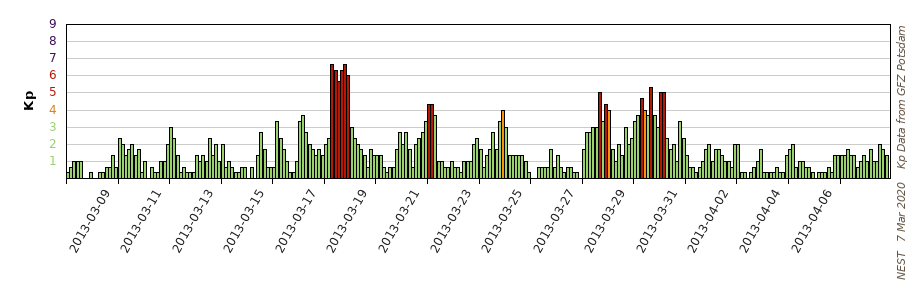}
    \caption{Variation of the 3-hour planetary index {$K_p$}. There are time periods for which the {$K_p$} changes by a large amount (from nest.nmdb.eu).}
    \label{fig:planetary-index}
\end{figure}
{Our simulations require the {$K_p$} index as input, which is measured every 3~hours. Therefore, we compute the cutoff rigidity assuming that {$K_p$} is constant within this time interval. The corresponding uncertainties is discussed in what follows.}  Fig.~\ref{fig:planetary-index} displays the variation of the 3-hour {$K_p$} index from March 7, 2013 to April 7, 2013. The figure shows that the {$K_p$}-index is highly variable and often changes by two from one to another interval. During extreme periods the {$K_p$} even increases up to a value of 6 corresponding to an IOPT of 7. In order to estimate the differences between quiet (IOPT = 1) and active (IOPT = 7) phases, computations of the cutoff rigidity were performed for two specific periods. In addition,  a period of an intermediate IOPT value of 4 has been investigated. Therefore, the conditions on March 15, 2013 (15:00~h, IOPT = 1), March 17, 2013 (12:00~h, IOPT = 7), and March 15, 2013 (03:00~h, IOPT = 4) at a mean \ac{ISS} flight altitude of 415~km have been modeled. 
\begin{figure}
    \centering
    \includegraphics[width=\columnwidth]{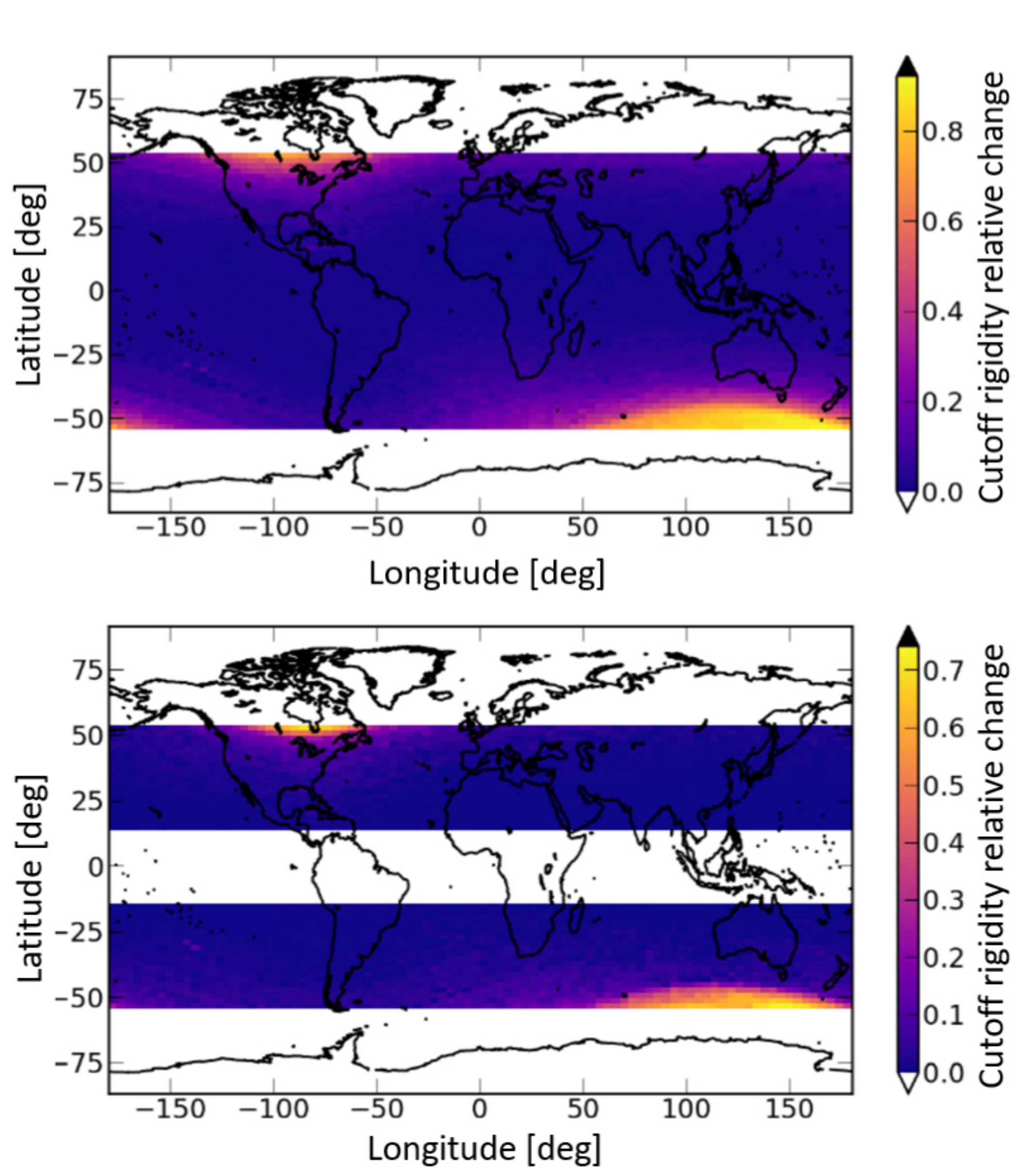}
    \caption{The two panel display the relative difference of the vertical cutoff rigidity maps on March 15, 2013 at 15:00 h {and March 17, 2013 at 12:00 h} are shown for IOPT=1 and 7 as well as  IOPT=1 and 4 {(on March 15, 2013 at 3:00 h)}, respectively. 
    We note that only regions of large relative differences are shown in the lower panel.}
    \label{fig:TSY89-IOPT-15}
\end{figure}
The upper panel of Fig.~\ref{fig:TSY89-IOPT-15} shows the relative differences  between the model results utilizing IOPT=7 and IOPT=1, while the lower panel displays the moderate differences between IOPT=1 and IOPT=4, respectively. The differences are growing towards the poles, reaching values around 0.9 and 0.7, respectively, for a few bins at the very highest latitudes considered. 
In particular, when the IOPT is higher (upper panel of Fig.~\ref{fig:TSY89-IOPT-15}) a high relative difference is also found at a certain distance from the highest latitudes. The comparison between the lowest and the extreme case (upper panel) as well as the comparison between the lowest and the moderate case can be utilized to estimate the uncertainty of the cutoff rigidity computations at a certain location. For most high-latitude locations it is lower than 20\% and 10\% for the extreme and the moderate case, respectively. Although the \ac{TSY89} model is not a good approximation of the external magnetic field under extreme conditions (IOPT above 3) we used the computation to estimate the maximum uncertainties. 
\subsection{Time corrections}
\begin{figure}
    \centering
    \includegraphics[width=\columnwidth]{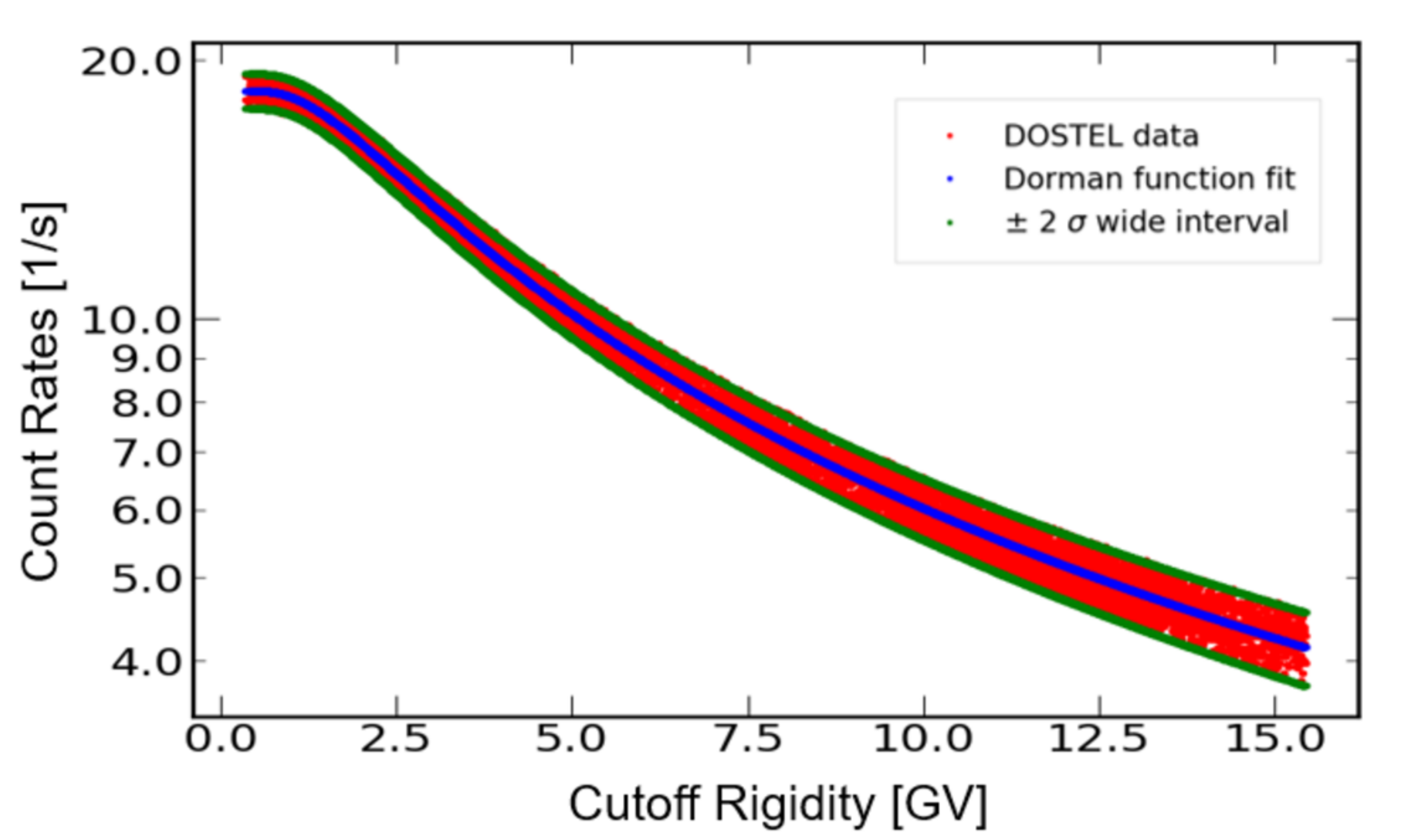}
    \caption{\ac{DOSTEL} count rates {(red dots)} as function of vertical cutoff rigidity for the time period from May 20, 2015 to June 7, 2015. The blue line is the result of the fit of Eq.~(\ref{eq:DORMAN}) to the data. The parameters found are $N_0 = 18.443 \pm 0.020$~1/s, $\alpha = 4.044 \pm 0.0017$~GV , and  $k = 1.013 \pm 0.001$ (see also Tab.~\ref{tab:Dorman-periods}). We note that the DOSTEL count rates have been obtained by filtering out data which were not within the 2 $\sigma$ uncertainty band (green lines), where $\sigma$ is calculated considering the counts disposed in a Poisson distribution.}
    \label{fig:fit-dorman}
\end{figure}

As shown by \citeauthor{Labrenz-etal-2013} \citeyear{Labrenz-etal-2013} the count as well as the dose rate measured by the \ac{DOSTEL} need to be corrected for timing issues. { In this study, the authors state:} ''The relation between count rate and $R_c$ can be plotted for every 6-hour data file, which covers four 90 minutes orbits of the \ac{ISS}. To do this, orbit data of the \ac{ISS} were used to get the corresponding location for each 100-second count rate interval.'' The $R_c$ values, computed on a 1$^\circ$x1$^\circ$ grid, were used to get the $R_c$ values of the according positions. With this approach, we plot the measured count rates against the vertical geomagnetic cutoff rigidity values and determine the best time shift in order to obtain a distribution of measurements shown in Fig.~\ref{fig:fit-dorman}. Here, the blue line gives the fit of Eq.~(\ref{eq:DORMAN}). The corresponding parameters are summarized in the figure caption and Tab.~\ref{tab:Dorman-periods} (Appendix A). 
\subsection{Computation of the Yield function \label{sec:yieldfunction}}
The method used to calculate the yield function is the one reported in \citeauthor{Caballero-Lopez-Moraal-2012} \citeyear{Caballero-Lopez-Moraal-2012}. In a first step, Eq.~(\ref{eq:yield-def}) can be re-written as
\begin{equation}
    \frac{\partial N (R, t, x)}{\partial R} = \sum\limits_{i} dR \; j_i(R) \; Y_i(R) \label{eq:3.5},
\end{equation}
where the term on the left reflects the differential count rate, which  can be approximated by the derivative of the Dorman function given in Eq.~(\ref{eq:DORMAN}) (see also \citeauthor{Dorman-etal-1970} \citeyear{Dorman-etal-1970}):
\begin{equation}\label{eq:Dorman-differential}
    \frac{d N(R_C, t)}{d R_C} = N_0 \cdot \alpha \cdot (k-1) \cdot R^k \exp(-\alpha \cdot R^{-k+1}).
\end{equation}
According to \citeauthor{Caballero-Lopez-Moraal-2012} \citeyear{Caballero-Lopez-Moraal-2012} the term on the right side of Eq.~(\ref{eq:3.5}) further  can be simplified by
\begin{equation}
    \frac{d N(R, t)}{d R} = j_H(R,t) \cdot Y_H(R) + j_{He} \cdot Y_{He} + j_{CNO} \cdot Y_{CNO} + j_{Other} \cdot Y_{Other}.
\end{equation}
The proton yield function $Y_H$ can be obtained as follows:
\begin{equation} \label{eq:yield-calc}
    Y_H(R) = \frac{\frac{d N(R, t)}{d R}}{j_H(R,t)  + j_{He}(R, t) \cdot \frac{Y_{He}(R)}{Y_H(R)} + j_{CNO}(R, t) \cdot \frac{Y_{CNO}(R)}{Y_H(R)} + j_{Other}(R, t) \cdot \frac{Y_{Other}(R)}{Y_H(R)}}
\end{equation}
\begin{table}[!t]
    \centering
    \begin{tabular}{c|c|c}
         Parameters & Neutron Monitor & Free space \\
         \hline
         $F_0$ & 2 & 2\\
         $P_0$ & 0.45 & 5.5\\
         $a$ & 1.4 & 1.4 \\
         $\gamma_1$ & 0 & 0 \\
         $\gamma_2$ & 10 & 0.4 \\
    \end{tabular}
    \caption{Comparison of the parameters of the function F for Neutron Monitor and in free space.}
    \label{tab:f_r}
\end{table}
\begin{figure}[htb]
    \centering
    \includegraphics[width=\columnwidth]{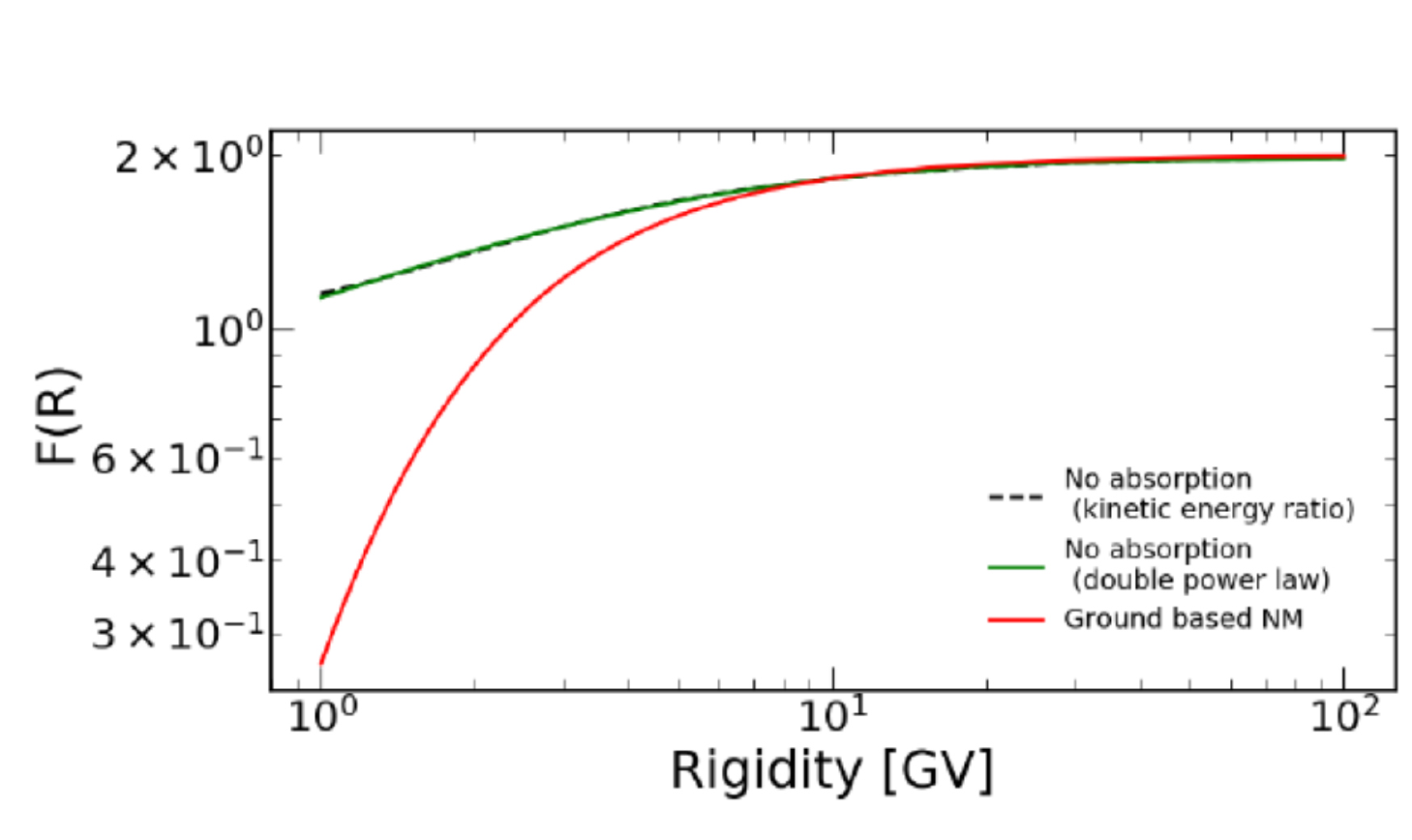}
    \caption{F(R) for no absorption condition (space-borne) represented by the kinetic energy ratio (dashed line) and a double power law (green line). In comparison F(R) of  ground based Neutron Monitors is shown as red line. The parameters for the two functions are listed in Tab.~\ref{tab:f_r}.}
    \label{fig:f_r}
\end{figure}
With the following approximations  
\begin{itemize}
    \item the relative abundance ratios at 10.6~GV were taken from \citeauthor{Gaisser-etal.-2016} \citeyear{Gaisser-etal.-2016} with 84.9~\%, 13.3~\%, 1.1~\% and 0.8~\% for Hydrogen, Helium, CNO and others, respectively,
    \item for heavy ions ($Z\geq4$) a $\frac{A}{Z}$ ratio of 2 has been assumed, and therefore $Y_{i} (R) =\frac{Z}{2} Y_{He} (R)$ 
    \item \citeauthor{Clem-Dorman-2000} \citeyear{Clem-Dorman-2000} found a relation $\frac{Y_{He}(R)}{Y_H(R)} = F(R)$ from Monte Carlo simulations that has been approximated by \citeauthor{Caballero-Lopez-Moraal-2012} \citeyear{Caballero-Lopez-Moraal-2012} in form of a double power law depending on five parameters (Eq.~\ref{eq:doule-power-law} given in Tab.~\ref{tab:f_r}).
\end{itemize} 
the denominator can be written as:
\begin{equation}
    j_H(R) + F (R) \cdot J_{He}(R,t) \cdot \left( 1+3.5\cdot \frac{J_{CNO}(R.t)}{J_{He}(R,t)} + 5.6 \cdot \frac{J_{Other}(R.t)}{J_{He}(R,t)}\right),  
\end{equation}
and can be reduced to
\begin{equation}\label{eq:yield-nenner}
    j_{H}(R)+1.584\cdot F(R) \cdot j_{He}(R,t)
\end{equation}
when the relative abundance ratios are implemented. Thus, the proton yield function $Y_H(P)$ can be calculated from the measured hydrogen and helium rigidity spectra and the cutoff rigidity dependence of the instrument if the shielding/shooting differences between hydrogen and helium $F(R)$ is known. In the case of neutron monitors this $F(R)$ was derived by \citeauthor{Clem-Dorman-2000} \citeyear{Clem-Dorman-2000}. The method developed by \cite{Caballero-Lopez-Moraal-2012} follows an analysis through NM data, for which highly shielding conditions are valid. Taking into account that the shielding of the ISS is not exactly known and, thus, wrongfully may be assumed to be nearly negligible, the real conditions could correspond to values of F(R) well within the case of neutron monitors (solid red line in Fig.~\ref{fig:f_r}) with a shielding of 1000 g/cm$^2$ and the case of no shielding (solid green line in Fig.~\ref{fig:f_r}). It shows that both cases differ significantly for rigidities below 3 GV. For more detail, Fig.~\ref{fig:free_NM_ratio} shows the ratios between the highest and the lowest yield functions of the sample of the four quiet periods given in Table \ref{tab:quiet-periods}. As can be seen, the shielding effect becomes negligible for particles with energies above 3~GV.

However, although these substantial differences occur an approximation by a double power-law can be found: 
\begin{equation}\label{eq:doule-power-law}
    F(R) = F_0(R_0^a + R^a)^{\frac{(\gamma_1-\gamma_2)}{a}}\cdot R^\gamma_2,
\end{equation}
with $\gamma_1$, $\gamma_2$, $R_0$, $F_0$ and $a$ as the two spectral indices, the roll over rigidity, and the ratio at $R=\infty$, respectively. The corresponding parameters are summarized in Tab.~\ref{tab:f_r}. 
\begin{figure}[!t]
    \centering
    \includegraphics[width=\columnwidth]{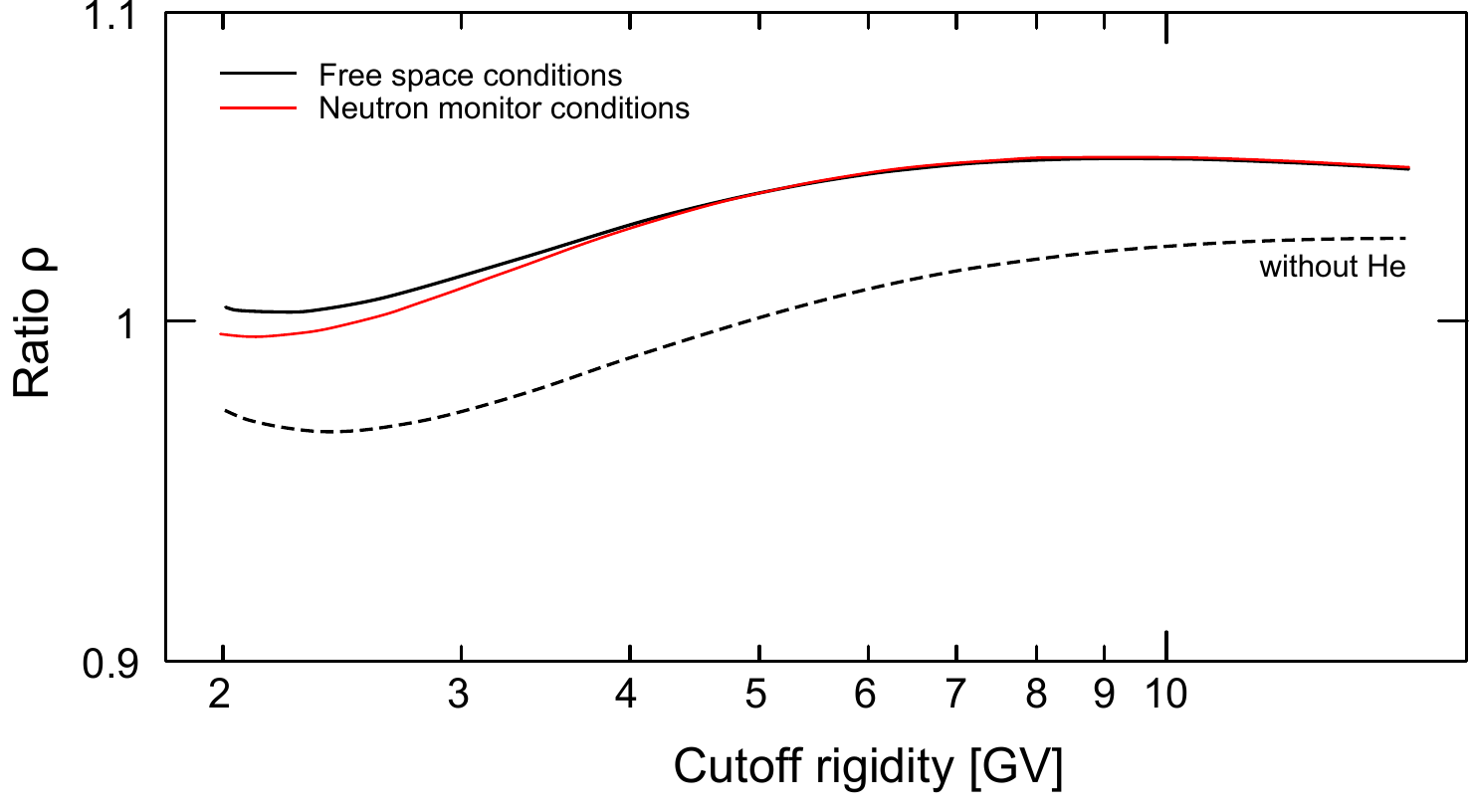}
    \caption{{Ratio between the highest and the lowest yield functions of the sample of the four quiet periods given in Table \ref{tab:quiet-periods}. The ratios are calculated based on Eq.~(\ref{eq:yield-calculation}) (solid lines). In addition,  the coefficient of F(R) is set to 0, therefore excluding the He contribution (dashed line). Here a direct comparison between free space conditions (black curves) and neutron monitor conditions (red curve) is shown.} 
    }
    \label{fig:free_NM_ratio}
\end{figure}
\section{Data analysis}
In order to determine the \ac{ISS} \ac{DOSTEL} specific yield function we utilized the following equation
\begin{equation}
    Y_H(R) = \frac{\frac{d N(R, t)}{d R}}{j_{H}(R)+1.584\cdot F(R) \cdot j_{He}(R,t)}
    \label{eq:yield-calculation}
\end{equation}
resulting from Eq.~(\ref{eq:yield-calc}) and Eq.~(\ref{eq:yield-nenner}). In order to determine the quantities on the right side of this equation   
we choose the following approach: 
\begin{table}[]
    \centering
    \begin{tabular}{|c|c|c|c|c|c|}
        \hline
        Period & Start time & End time &  hours & percentage & modulation \\
         & & & quiet times & quiet times & parameter\\
         & & &  &  & in MV \\
         \hline
         1 & 2/3/2014 & 21/3/2014 & 435 & 92 & 652 \\
         & 3:00 UTC & 12:00 UTC & & & \\
         \hline
         2 & 20/5/2015 & 7/6/2015 & 435 & 97 & 660.5\\
         & 6:00 UTC & 15:00 UTC & & & \\
         \hline
         3 & 8/9/2016 & 24/9/2016 & 357 & 90 & 436\\
         & 12:00 UTC & 21:00 UTC & & & \\
         \hline
         4 & 26/4/2017 & 15/5/2017 & 414 & 95 & 363\\
         & 3:00 UTC & 6:00 UTC & & & \\ \hline
    \end{tabular}
    \caption{Four quiet time periods associated to long term low IOPT values (IOPT$ < $4). The total hours and the relative contributions are quantified in the  fourth and fifth column. The last column gives the modulation parameter $\phi$ from \citeauthor{Usoskin-etal-2005} \citeyear{Usoskin-etal-2005}.}
    \label{tab:quiet-periods}
\end{table}
\begin{description}
    \item[$\frac{dN}{dR}$:] We selected four time periods that include a large number of days for which the planetary index {$K_p$} and, therefore, the IOPT is below~3. The start and end times of the selected periods and other corresponding quantities are summarized in Tab.~\ref{tab:quiet-periods}. Note, that the \ac{ISS} was for all periods in the +XVV configuration with one exception during the first period. A maneuver oriented the station to +ZVV on 10.03.2014 23:10 till 11.03.2014 00:11 then the station returned to +XVV. Since the time during the different orientation was short compared to the full period we neglect the effect in what follows. However, the derived yield functions will therefore only be valid during the +XVV configuration. 
\begin{enumerate}
    \item During the second of the four periods (on May 20, 2015), a \ac{SEP} event was registered by the \ac{EPHIN} aboard \ac{SOHO}. As {a} comparison, the period-dependent 30~minute averaged count rate variations of above 50~MeV protons are shown in the panels of Fig.~\ref{fig:SOHO-FLUX}. Note that time profiles indicate variations of several percent{s} during each period at rigidities below 2~GV \cite{Kuehl-etal-2015}.
    \item  The approximation of the rigidity-dependent count rate profiles of the four periods utilizing the Dorman function given in Eq.~(\ref{eq:DORMAN}) are displayed in the panels of Fig.~\ref{fig:2020-03-16-Dorman}. Tab.~\ref{tab:Dorman-periods} summarizes the fit parameter and their uncertainties, while Fig.~\ref{fig:2020-03-16-differential-Dorman} shows the  differential spectra according to Eq.~(\ref{eq:Dorman-differential}).
\end{enumerate}
\begin{figure}
    \centering
    \includegraphics[width=\columnwidth]{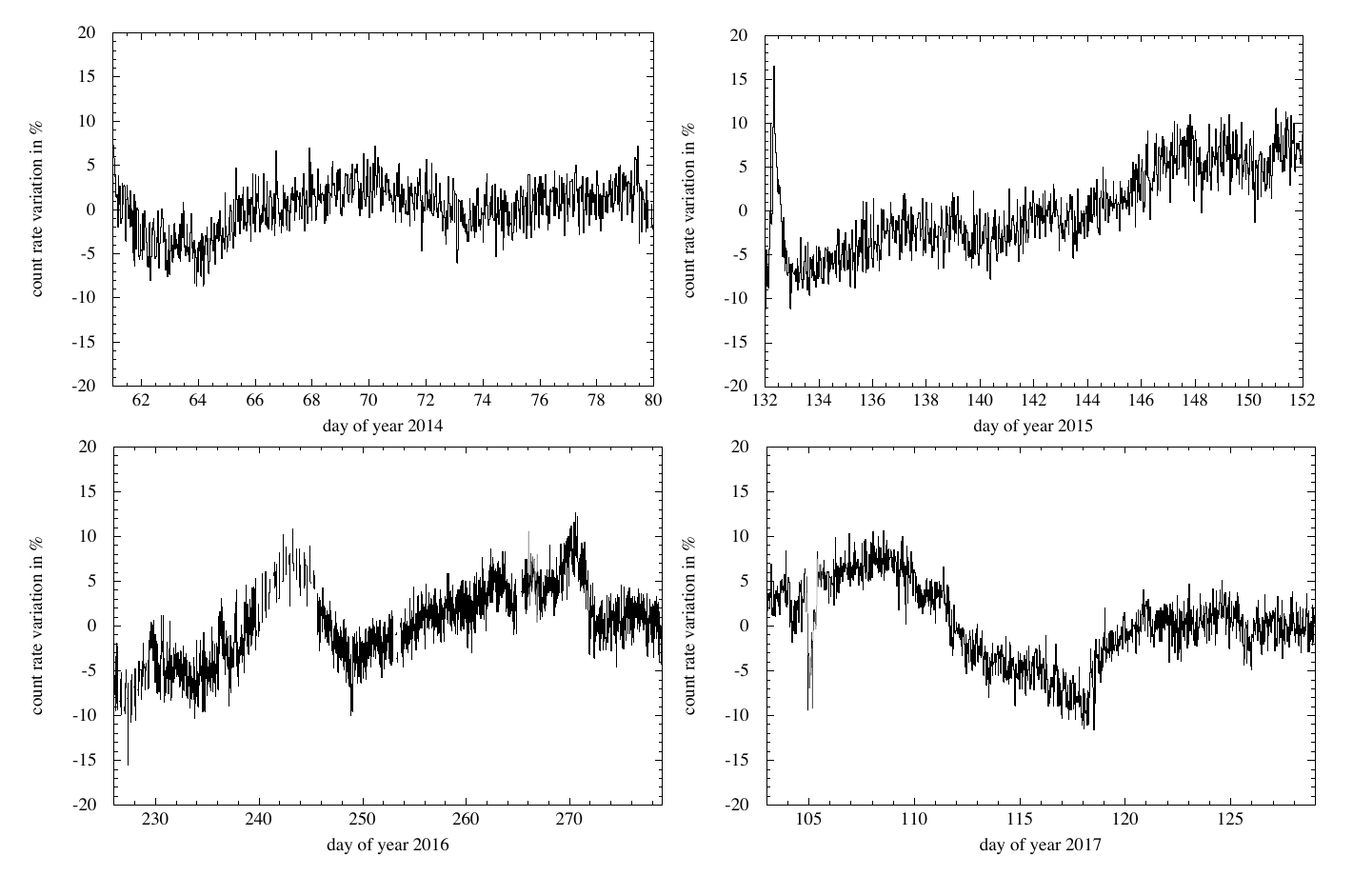}
\caption{Count rates of the above 50 MeV/nucleon proton and helium channel (black curve) of the \ac{EPHIN} on board \ac{SOHO} for the four periods given in Tab.~\ref{tab:quiet-periods}.} 
    \label{fig:SOHO-FLUX}
\end{figure}
    \item[$j_H, j_{He}:$] The primary proton and Helium fluxes measured with \ac{AMS}  are shown in the left and right panel of Fig.~\ref{fig:ams-spectra}, respectively. This experiment on-board the \ac{ISS} measures the rigidity spectra for different ions in the range of a few GV up to several 100 TV. The relative precision of the fluxes is below $\sim 5$~\%. The lower limit of the instrument is 1~GV for protons and 2~GV for Helium \cite{Aguilar-etal-2018}.  The data were obtained from https://tools.ssdc.asi.it/CosmicRays/ and are given as 27~day averages up to 2017. Here, we utilized the files corresponding to the selected periods in Tab.~\ref{tab:quiet-periods}. A double power-law was used to fit both proton and Helium spectra. The parameters and their uncertainties are summarized in Tab.~\ref{tab:fit-prptpn-helium}. 
\begin{figure}[tb]
    \centering
    \includegraphics[width=\columnwidth]{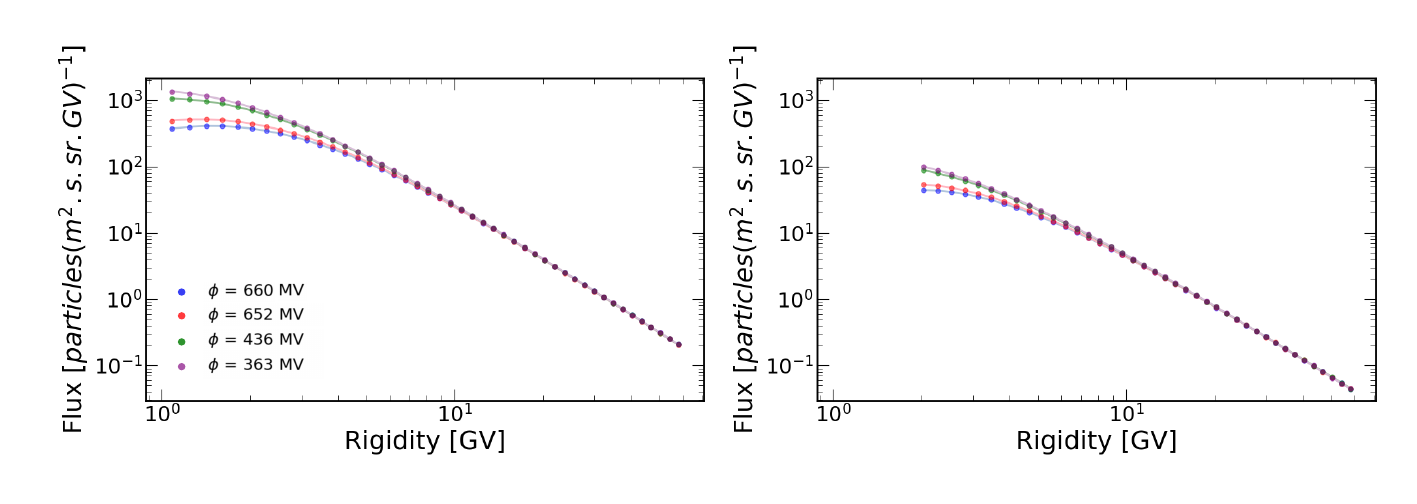}
    \caption{Energy spectra of protons (left panel) and helium (right panel) from \ac{AMS} during the  periods in Tab.~\ref{tab:quiet-periods}. Different color-coding corresponds to periods of different solar activity. The corresponding modulation parameters $\phi$ have been taken from \url{http://cosmicrays.oulu.fi/phi/Phi_mon.txt}.}
    \label{fig:ams-spectra}
\end{figure}
\item[$F(R):$] The method developed by \citeauthor{Caballero-Lopez-Moraal-2012} \citeyear{Caballero-Lopez-Moraal-2012} follows an analysis through NM data, for which high shielding conditions are valid. Bulk shielding of the ISS is substantial \cite{Stoffle-etal-2012,Doynde-etal-2019} but is approximately an order of magnitude less than that of neutron monitors. The effect of shielding on the yield function is important for rigidities less than 3~GV but of little importance at higher rigidity. Fig.~\ref{fig:f_r} shows F(R) for free space (green curve) and for typical neutron monitor shielding of 1000~g~cm$^{-2}$ (red curve). The cases are identical above 7~GV and differ only slightly between 3 and 7~GV. ISS shielding falls in between the two curves, closer to the free space case. For more detail, Fig.~\ref{fig:free_NM_ratio} shows the ratios between the highest and the lowest yield functions of the sample of the four quiet periods given in Tab.~\ref{tab:Dorman-periods}. As can be seen, the shielding effect becomes negligible for particles with energies above 3~GV. The yield functions for both scenarios have been computed, and their differences are taken into account when computing the uncertainties. 
\end{description}
\begin{figure}
    \centering
    \includegraphics[width=\columnwidth]{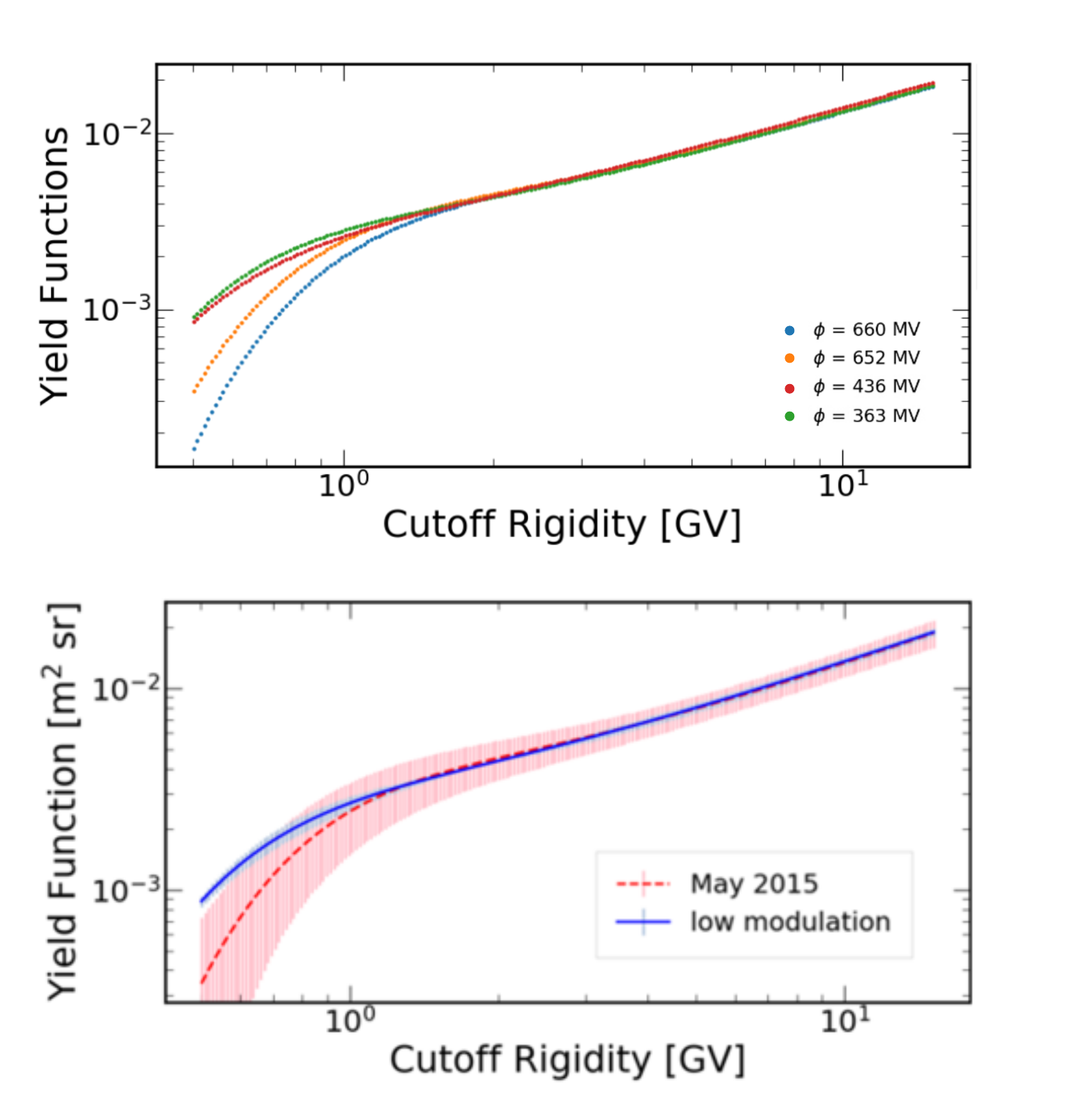}
    \caption[Yield functions]{{Upper panel: Yield functions for the four different modulation phases shown in Fig.~\ref{fig:ams-spectra} between 660 MV (in blue) and 363 MV (in green). Lower panel: }The blue and red curve display the results of our analysis for the two periods at solar minimum in 2016 and 2017 (in blue) and during intermediate activity in 2015  (in red), respectively. We find large deviation below 1~GV rigidity (for details see text).
    }
    \label{fig:comparision-yields}
\end{figure}
Fig. ~\ref{fig:comparision-yields} displays the results for solar minimum conditions at the declining phase of solar cycle 24 in 2015. All yield functions agree very well for rigidities above 1~GV, however showing significant deviations at cutoff rigidities below 0.6~GV. It is important to note that the error bands for the case of low modulation are calculated as the differences between the yield functions for each of the two periods, while in the case of high solar activity, the error band is calculated by propagating the uncertainties on the parameters resulting from the previous fits of the count rates and flux data. The increase of the latter uncertainty with decreasing rigidity is due to the fact that the effect of the shielding is estimated by the extreme cases as detailed above. At rigidities below 2 GV the error is therefore largely overestimated. 
{
The second period in 2015 has been used to validate our approach by calculating the count rate rigidity profile from the yield function from the low modulation periods and the measured proton and helium spectra by \ac{AMS}. From eq.~\ref{eq:yield-calculation} we get:
\begin{equation}
    N(R,t) = \int_{R_c}^\infty \frac{dN (R,t)}{dR} dR = \int_{R_c}^\infty Y_H(R) \cdot \left(j_{H}(R)+1.584\cdot F(R) \cdot j_{He}(R,t) \right) dR \label{eq:count-calculation}
\end{equation}
with $Y_H(R)$ determined at solar minimum. The result is shown in Fig.~\ref{fig:count-calculation}. }
\begin{figure}
    \centering
    \includegraphics[width=\columnwidth]{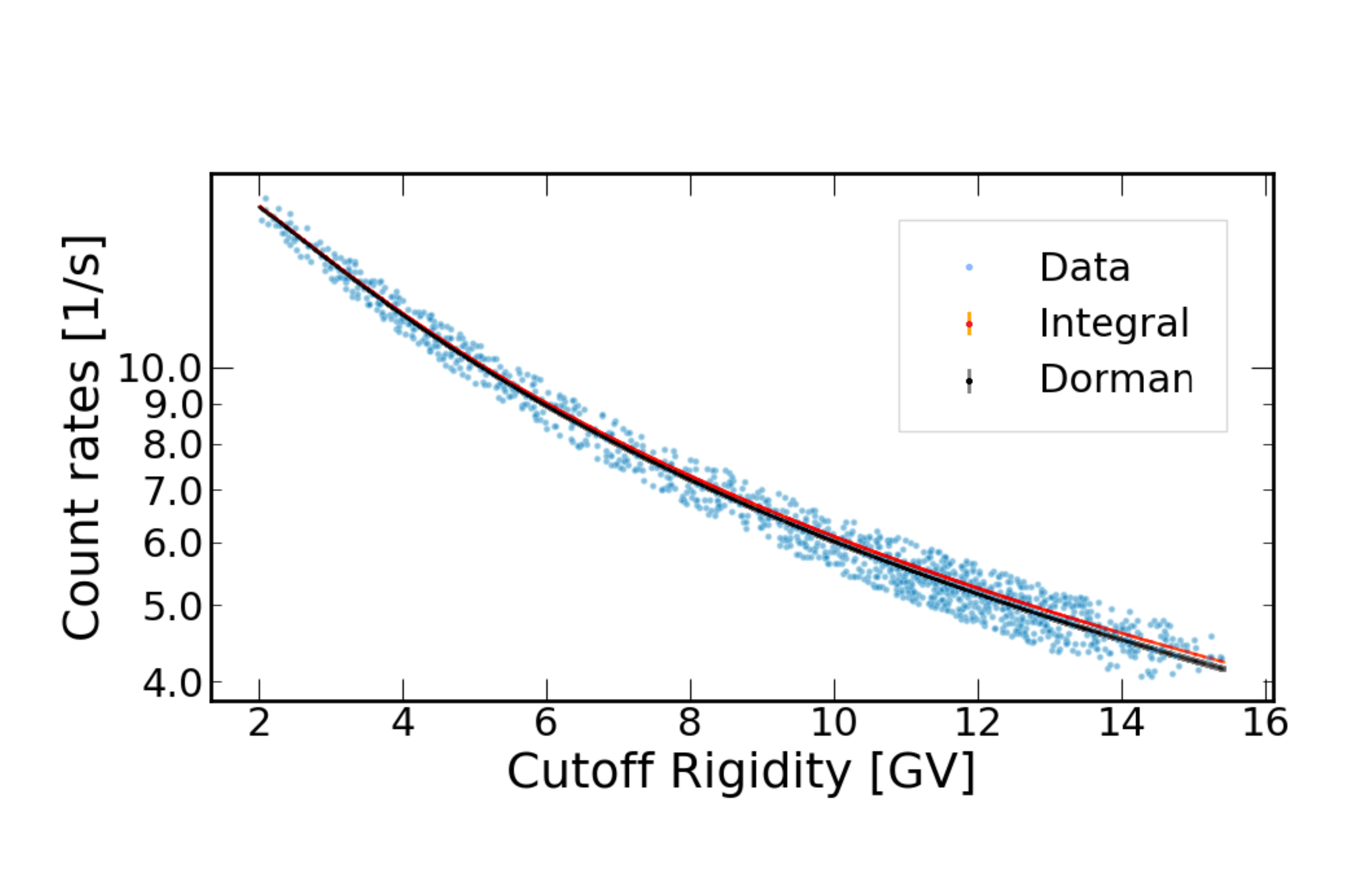}
    \caption{Measured count rates of the first period (blue dots). In addition, the approximation by the Dorman function (green line) and the Integral function derived by the yield function (Eq.~(\ref{eq:count-calculation}), red line) are shown.}
    \label{fig:count-calculation}
\end{figure}
\subsection{Yield function for the dose rate}
\begin{figure}
    \centering
    \includegraphics[width=\columnwidth]{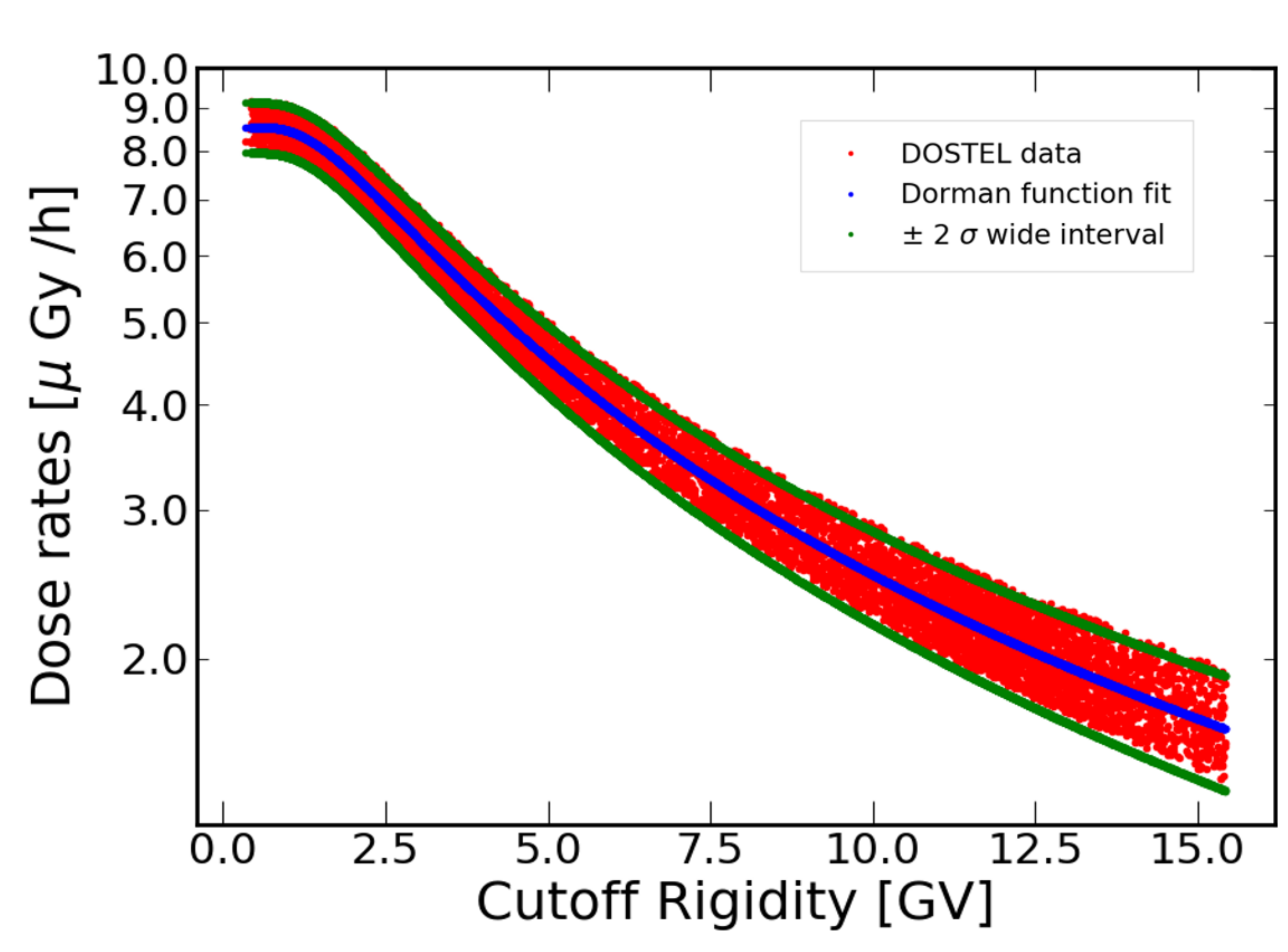}
    \caption{Dose rate measurement (red dots) and fit of the Dorman function to the data set for the second period. The resulting parameters are $N_0 = 8.513 \pm 0.015$~1/s, $\alpha = 4.480 \pm 0.031$~GV and, $k = 1.115 \pm 0.002$}
    \label{fig:dr-dorman-per2}
\end{figure}

\begin{figure}
    \centering
        \includegraphics[width=\columnwidth]{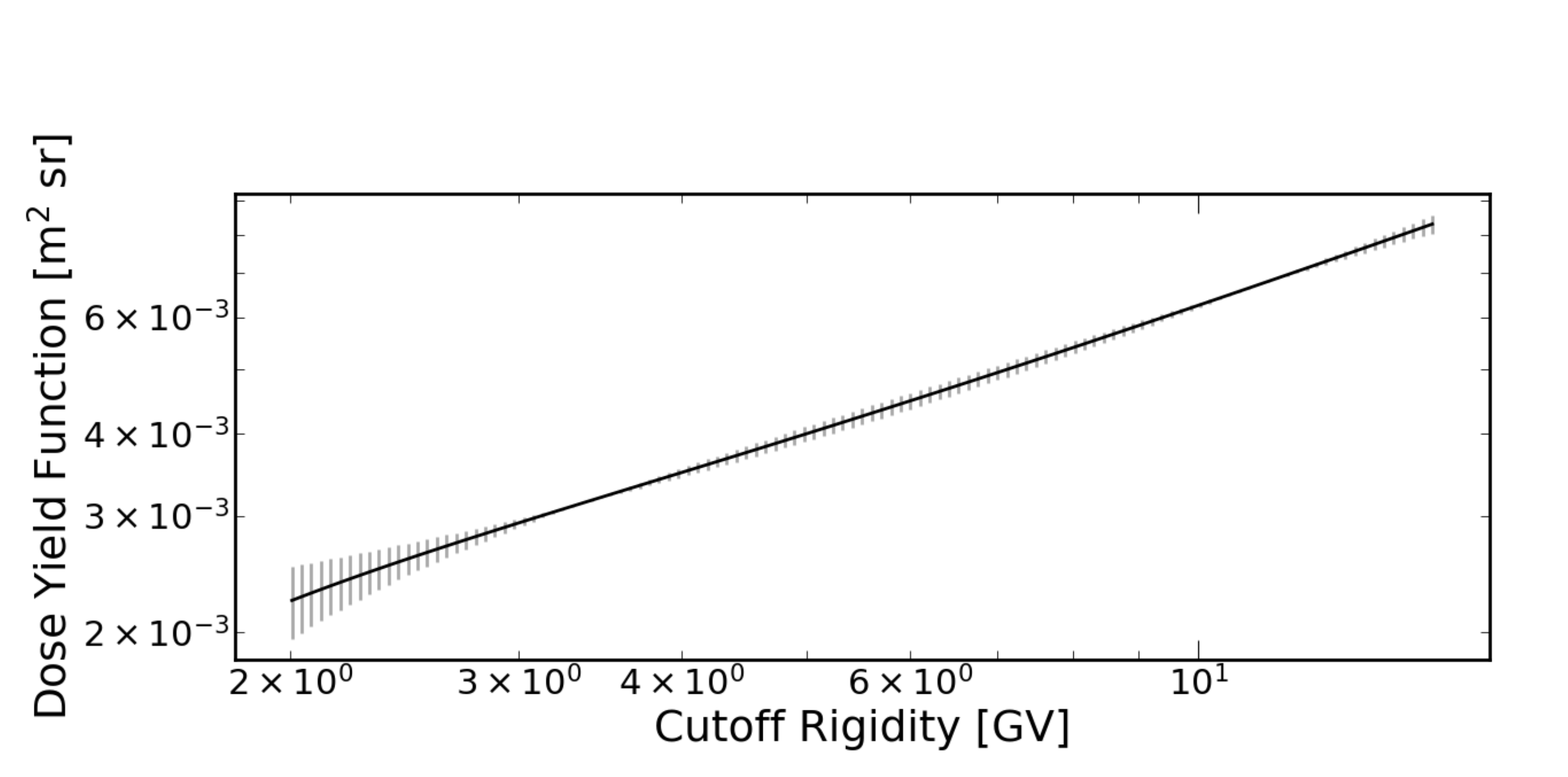}
    \caption{Average yield function calculated from the dose rates, valid for rigidities above 2~GV and shown by the black line.}
    \label{fig:Yield-Dose-Rate}
\end{figure}

The method used to calculate the count rate yield function was applied to the dose rate measurements of \ac{DOSTEL}. Therefore, the measured dose rates were {fitted} with the Dorman function for the four periods. The result are summarized in Tab.~\ref{tab:Dorman-periods}. With $\chi^2$-values between 1 and 1.5 for all the four sets the theoretical values are in good agreement with the measurements. An example is shown in Fig.~\ref{fig:dr-dorman-per2}, where the calculated Dorman function is plotted together with the corresponding dose rate data sample.
Following the discussion on the count rate yield function, we applied the same procedure including the uncertainties for the function $F(R)$ to the dose rates. However, due to the significant uncertainties below 2 GV, we decided to provide the averaged yield for the dose rate for vertical cutoff rigidities above 2~GV only as shown in Fig.~\ref{fig:Yield-Dose-Rate}. 
\section{Summary and Conclusion \label{sec:summary-conclusion}}
{In this study we determined the yield function for the count and dose rate of the \ac{DOSTEL} that is part of the DOSIS~3D experiment on board the \ac{ISS}. Our analysis is based on the method suggested by \citeauthor{Caballero-Lopez-Moraal-2012} \citeyear{Caballero-Lopez-Moraal-2012}. In order to compute this function, the initial count and dose rate data have to be corrected for a time offset to relate them to cutoff rigidity values. In order to achieve this we follow \citeauthor{Labrenz-etal-2015} \citeyear{Labrenz-etal-2015} and calculated time- and space-dependent cutoff rigidities utilizing PLANETOCOSMICS \citeauthor{Desorgher-etal-2009} \citeyear{Desorgher-etal-2009}. To reduce the computing time, simulations were performed once for three hour intervals. During each period, we used the mean altitude of the \ac{ISS} and magnetic field disturbance level IOPT. The Earth's magnetic field is modeled by the \ac{IGRF} and the \ac{TSY89} model. This model has been shown  to describe the Earth's magnetic field during low disturbance levels quiet well. The analysis was therefore restricted to periods low planetary index \textit{K}\textsubscript{p} ($< 3$). Four time-periods of $\simeq{15 - 20}$ days between the years 2014 and 2017 were selected. An inspection of the proton flux at about 1~GV during these periods was performed. The first and second periods (March 2014, May 2015) were influenced by remains of \ac{SEP} events, which affected the result for these periods. Following the approach by \citeauthor{Labrenz-etal-2015} \citeyear{Labrenz-etal-2015} the \ac{DOSTEL} files were corrected for different time offsets. For this step, we computed from the \ac{ISS} location in time the corresponding cutoff rigidities. 

Following the steps described in \citeauthor{Caballero-Lopez-Moraal-2012} \citeyear{Caballero-Lopez-Moraal-2012}, the yield function of \ac{DOSTEL} was computed. For this purpose 
\begin{enumerate}
    \item the count and dose rate rigidity profiles were fitted by the Dorman function eq.:~\ref{eq:DORMAN}.
    \item the proton and helium fluxes were obtained from \ac{AMS} - 02 for the time periods in question. The rigidity spectra were fitted by a double power-law. 
    \item the function $F(R)$ is required, where $R$ indicates rigidity. This function represents the computed ratio between the yield function for protons and Helium. In our analysis $F(R)$ for free-space conditions that means no absorption by the \ac{ISS} to the incoming particles was approximated by a double power-law.
\end{enumerate} 
The yield functions were calculated for the periods in the declining and solar minimum phase of the solar cycle for rigidities above 2 GV only because the \ac{AMS} helium spectra are only available above 2 GV. Due to an  extrapolation to lower rigidities and due to the complexity of the model, the determination of the function at lower rigidities is not reliable, and the fit parameters get highly correlated. In addition, the uncertainties of all three steps above were taken into account, leading to large errors of the yield functions below 2~GV. The count rates of the first and second period were computed to validate the method, using the \ac{AMS} spectra. The comparison of the model with the measured count rates leads to a reasonable agreement, indicating that the description by the yield function can be used to compute the count rate variation for different levels of solar activity. To improve our approach, further periods need to be investigated and primary rigidity spectra should become available below 2~GV for helium and below 1~GV for protons. }

\acknowledgments
We acknowledge the support from  Christian-Albrechts-Universitaet (CAU) zu Kiel and Università degli Studi di Milano. K\textsubscript{p} measured values are provided by Helmholtz Centre Potsdam (ftp://ftp.gfz-potsdam.de/pub/home/obs/kp-ap/tab/, GFZ German Research Centre for Geosciences). \ac{AMS} flux data are provided by the Space Science Database Centre via the Cosmic Ray Database (https://tools.ssdc.asi.it/CosmicRays/). \ac{SOHO} \ac{EPHIN} and \ac{DOSTEL} measurements can be obtained from http://ulysses.physik.uni-kiel.de/costep/ and http://ulysses.physik.uni-kiel.de/exchange/publications/DOSTEL/, respectively. We acknowledge the NMDB database (www.nmdb.eu), founded under the European Union's FP7 programme (contract no. 213007) for providing data. BH and KH received funding from the European Union’s Horizon 2020 research and innovation programme under grant agreement No 870405. BH and KH further acknowledge the International Space Science Institute and the supported International Team 441: High EneRgy sOlar partICle EventsAnalysis (HEROIC) {and Team 464: The  Role  Of  Solar  And  Stellar  Energetic  Particles  On (Exo)Planetary Habitability (ETERNAL)}.

\bibliography{2021-02-11}

\appendix
\section{{$K_p$} and IOPT relation}
\begin{table}[htb]
\begin{tabular} {|c|c|c|c|c|c|c|c|}
\hline
{$K_p$} & 0,0+ & 1-,1-1+ & 2-,2,2+ & 3-,3-3+ & 4-,4,4+ & 5-,5-5+ & $> 6-$ \\ 
\hline
$IOPT$ & 1 & 2 & 3 & 4 & 5 & 6 & 7 \\
\hline
\end{tabular}
\caption{Table showing the correspondence between IOPT and {$K_p$}.}
\label{tab:KP-IOPT}
\end{table} 
~\clearpage
\section{Parameter of the Dorman function}
\begin{table}[htb]
    \centering
\begin{tabular} {|c|c|c|}
\hline
 & Count rate Fit parameters & $\chi$\textsuperscript{2} \\
\hline
1 &  N\textsubscript{0} = 16.674  $\pm$ 0.017 1/s & 1.044\\& $\alpha$ = 4.592  $\pm$ 0.019  GV\textsuperscript{k}&\\ & $k$  = 1.022  $\pm$  0.001& \\
\hline
2 &  N\textsubscript{0} = 18.443 $\pm$ 0.020 1/s & 1.051\\ & $\alpha$ = 4.044 $\pm$ 0.017 GV\textsuperscript{k} &\\ & $k$  = 1.013  $\pm$  0.001 &\\
\hline
3 &  N\textsubscript{0} = 22.944  $\pm$ 0.030 1/s & 1.068\\& $\alpha$ = 3.062 $\pm$ 0.012 GV\textsuperscript{k} &\\ & $k$  = 1.002  $\pm$  0.001 & \\
\hline
4 &  N\textsubscript{0} = 24.618  $\pm$ 0.033 1/s & 1.085\\& $\alpha$ = 2.861 $\pm$ 0.001  GV\textsuperscript{k}&\\ & $k$  = 0.991  $\pm$  0.001&\\
\hline 
\end{tabular}
\begin{tabular} {|c|c|c|}
\hline
 & Dose rate fit parameters & $\chi$\textsuperscript{2} \\
\hline
1 &  N\textsubscript{0} = 7.677  $\pm$ 0.013 1/s & 1.150\\& $\alpha$ = 5.081  $\pm$ 0.036  GV\textsuperscript{k}&\\ & $k$  = 1.129  $\pm$  0.003& \\
\hline
2 &  N\textsubscript{0} = 8.513 $\pm$ 0.015 1/s & 1.158\\ & $\alpha$ = 4.480 $\pm$ 0.031 GV\textsuperscript{k} &\\ & $k$  =  1.115  $\pm$  0.002 &\\
\hline
3 &  N\textsubscript{0} = 10.765  $\pm$ 0.023 1/s & 1.181\\& $\alpha$ = 3.455 $\pm$ 0.023 GV\textsuperscript{k} &\\ & $k$  = 1.120  $\pm$  0.002 & \\
\hline
4 &  N\textsubscript{0} = 11.711  $\pm$ 0.025 1/s & 1.217\\& $\alpha$ = 3.039 $\pm$ 0.019  GV\textsuperscript{k}&\\ & $k$  = 1.090  $\pm$  0.002&\\
\hline 
\end{tabular}

    \caption{Parameters for the Dorman function during period 1 (March 2 to March 21, 2014), 2 (May 20, to Jun 6, 2915), 3 (September 8 to September 24, 2016) and 4 (April 26 to May 14, 2017, respectively}
    \label{tab:Dorman-periods}
\end{table}
\begin{figure}[htb]
    \centering
    \includegraphics[width=\columnwidth]{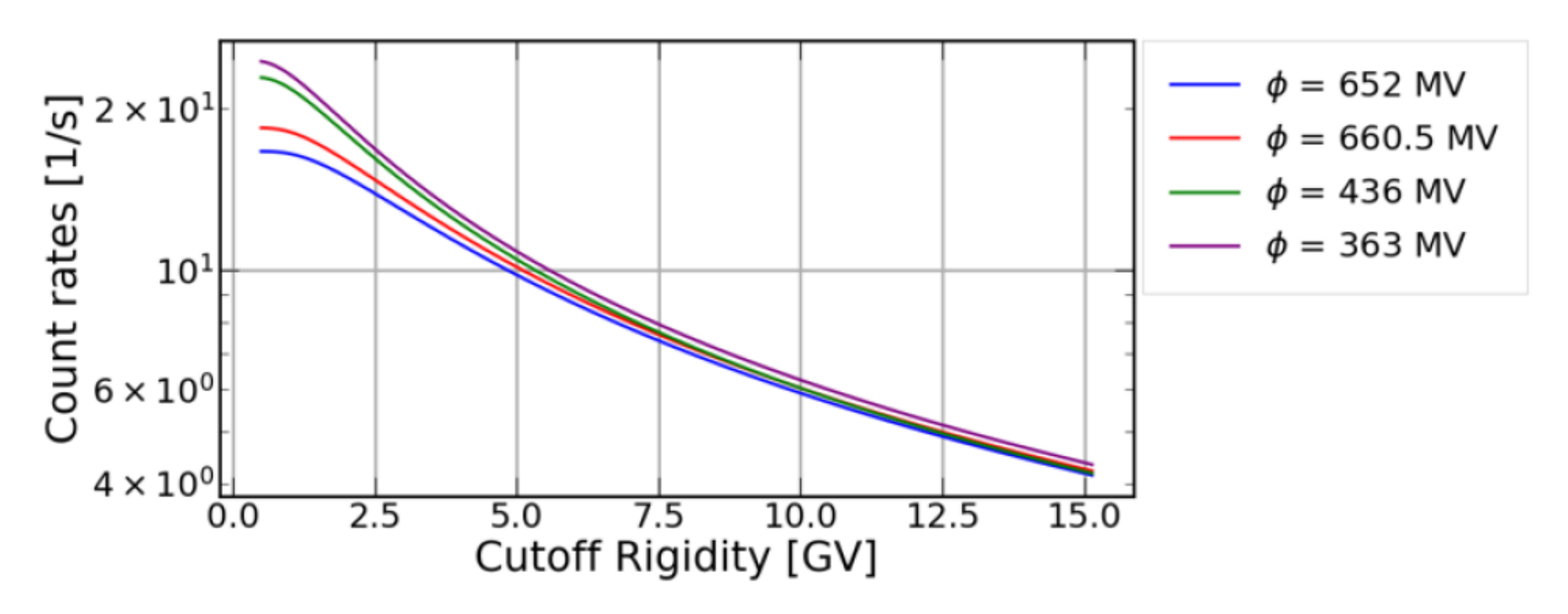}
    \caption{Approximation of the averaged count rates measured during the four periods analyzed in this paper.}
    \label{fig:2020-03-16-Dorman}
\end{figure}

\begin{figure}
    \centering
    \includegraphics[width=\columnwidth]{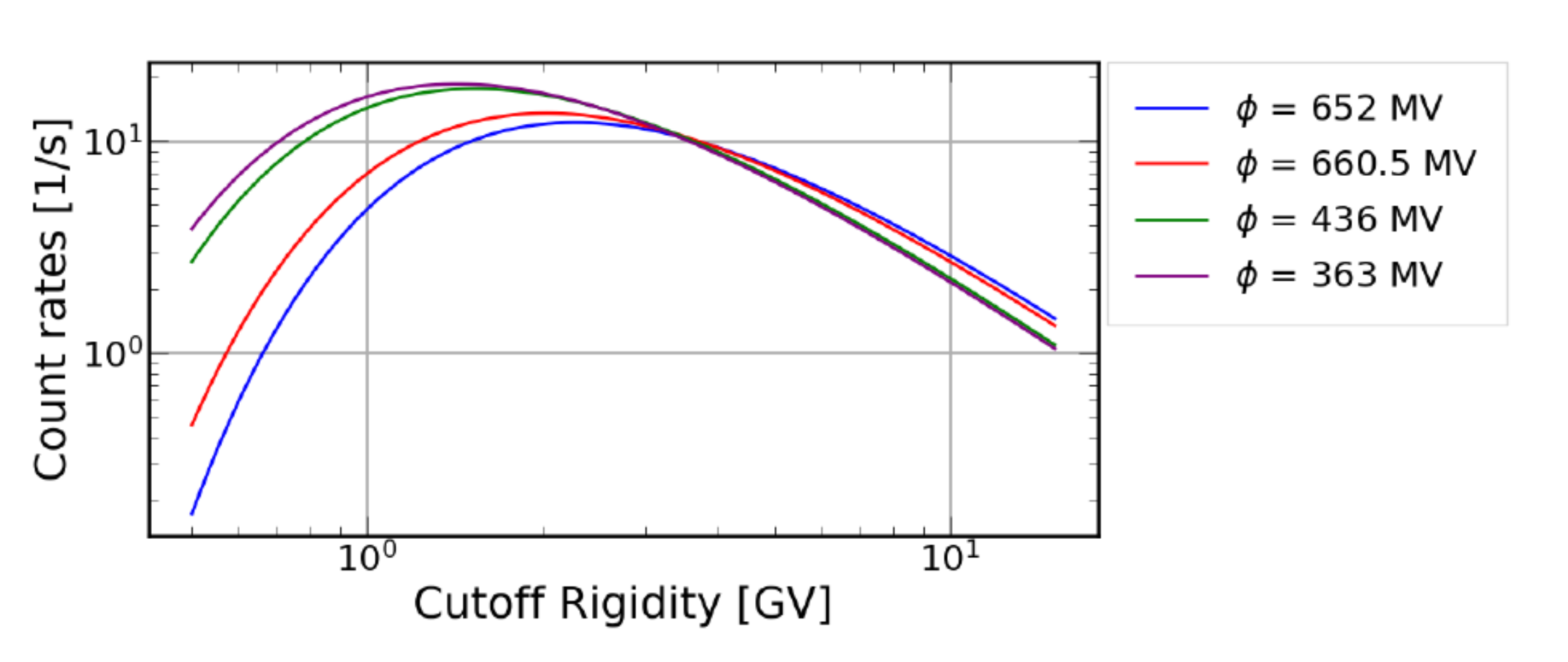}
    \caption{Differential count rate in the four periods analysed (with rigidity $P > 2$~GV). Corresponding values of the modulation parameter $\phi$ are shown.}
    \label{fig:2020-03-16-differential-Dorman}
\end{figure}
\clearpage
\section{Fit parameter for rigidity spectra}

\begin{table}[htb]
\begin{center}
\begin{tabular} {|c|c|c|}
\hline
 & Fit parameters (Proton flux) & Fit parameters (Helium flux) \\
\hline
1 &  F\textsubscript{0} = 24685.339	$\pm$	563.407 & F\textsubscript{0} = 4084.205 $\pm$  97.040 \\&  P\textsubscript{0} = 1.719 $\pm$ 0.074	 & P\textsubscript{0} = 0.600 $\pm$ 	0.029 \\ & a = 1.379 $\pm$ 0.023 & a = 1.192 $\pm$ 0.011 \\ & $\gamma$\textsubscript{1} = -2.872 $\pm$ 0.005 & $\gamma$\textsubscript{1} = -2.799 $\pm$ 0.006 \\ & $\gamma$\textsubscript{2} = 2.244 $\pm$ 0.165 & $\gamma$\textsubscript{2} = 11.653 $\pm$ 0.705 \\
\hline
2 &  F\textsubscript{0} = 23158.911 $\pm$	638.625	& F\textsubscript{0} = 3422.218	$\pm$  132.098 \\ & P\textsubscript{0} = 1.569	$\pm$ 0.100	 & P\textsubscript{0} = 0.800 $\pm$  0.119  \\ & a = 1.407 $\pm$ 0.032 & a = 1.296 $\pm$  0.036 \\ & $\gamma$\textsubscript{1} = -2.859 $\pm$ 0.007	& $\gamma$\textsubscript{1} = -2.760 $\pm$  0.009   \\ & $\gamma$\textsubscript{2} = 2.288 $\pm$ 0.256 & $\gamma$\textsubscript{2} = 8.144 $\pm$ 1.404  \\
\hline
3 &  F\textsubscript{0} = 22178.429	$\pm$ 407.866	 & 3145.324 $\pm$ 91.825   \\ & P\textsubscript{0} = 1.101	$\pm$ 0.087	 & P\textsubscript{0} = 0.800 $\pm$ 0.011   \\ & a = 1.360	$\pm$ 0.028 &  a = 1.313 $\pm$ 0.022  \\ & $\gamma$\textsubscript{1} = -2.850 $\pm$ 0.004  & $\gamma$\textsubscript{1} = -2.738 $\pm$ 0.007	  \\ & $\gamma$\textsubscript{2} = 2.626 $\pm$ 0.351 & $\gamma$\textsubscript{2} = 5.643	$\pm$ 0.217  \\
\hline
4 &  F\textsubscript{0} = 23344.391 $\pm$  789.663	 &  F\textsubscript{0} = 3502.970 $\pm$ 132.089  \\ &   P\textsubscript{0} = 0.860 $\pm$  0.160  &  P\textsubscript{0} = 0.800 $\pm$ 0.090  \\ & a = 1.247	$\pm$  0.047	 & a = 1.122 $\pm$ 0.027  \\ & $\gamma$\textsubscript{1} = -2.858 $\pm$ 0.008  & $\gamma$\textsubscript{1} = -2.766 $\pm$ 0.009  \\ & $\gamma$\textsubscript{2} = 2.967 $\pm$ 0.842 & $\gamma$\textsubscript{2} = 4.214 $\pm$ 0.775  \\
\hline 
\end{tabular}
\end{center}
\caption{Fit parameters that have been obtained for the proton and helium spectra during the four periods to approximate the \ac{AMS} rigidity spectra with eq.~\ref{eq:doule-power-law}}
    \label{tab:fit-prptpn-helium}
\end{table}

\end{document}